\documentclass[aps,prd,showpacs,nobibnotes,twocolumn,preprintnumbers,
nobalancelastpage,superscriptaddress]{revtex4}
\usepackage{latexsym}
\usepackage{dcolumn}% Align table columns on decimal point
\usepackage{bm}% bold math
\usepackage{natbib}
\input{epsf}
\newcommand{\be}{\begin{equation}}
\newcommand{\ee}{\end{equation}}
\newcommand{\ba}{\begin{eqnarray}}
\newcommand{\ea}{\end{eqnarray}}
\newcommand{\bi}{\begin{itemize}}
\newcommand{\ei}{\end{itemize}}
\newcommand{\bfi}{\begin{figure}[t]
\epsfxsize=9cm
\epsffile}
\newcommand{\efi}{\end{figure}}
\newcommand{\no}{\nonumber}

\begin{document}
\title{Peculiar Velocity Decomposition, Redshift Space Distortion and Velocity
  Reconstruction in Redshift Surveys -- I. The Methodology}    
\author{Pengjie Zhang}
\email[Email me at: ]{pjzhang@shao.ac.cn}
\affiliation{Key Laboratory for Research in Galaxies and Cosmology, Shanghai       
  Astronomical Observatory, Nandan Road 80, Shanghai, 200030,
  China}
\author{Jun Pan}
\affiliation{National Astronomical Observatories, Chinese Academy of Sciences,
20A Datun Rd., Chaoyang District, Beijing 100012, P. R. China}
\affiliation{Purple Mountain Observatory, 2 West Beijing Rd., Nanjing 210008, P. R.
China}
\author{Yi Zheng}
\affiliation{Key Laboratory for Research in Galaxies and Cosmology, Shanghai       
  Astronomical Observatory, Nandan Road 80, Shanghai, 200030,
  China}
\begin{abstract}
Massive spectroscopic surveys will  measure the redshift space distortion
(RSD) induced by galaxy peculiar velocity   to unprecedented accuracy and
open a new era of precision RSD cosmology.  We develop a new
method to improve the RSD modeling and to carry out robust
reconstruction of the 3D large scale peculiar velocity through galaxy redshift
surveys, in light of RSD. (1) We propose a mathematically unique and physically
motivated decomposition of peculiar velocity into three eigen-components: an
irrotational component completely correlated with the underlying density
field (${\bf v}_\delta$), an irrotational component uncorrelated with the density field
(${\bf v}_S$) and a  rotational (curl) component (${\bf v}_B$). The three
components have different origins, different scale dependences and different
impacts on RSD. (2)
This decomposition has the potential to simplify and improve the RSD modeling. (I)
${\bf v}_B$ damps the redshift space clustering.  (II) ${\bf v}_S$
causes both damping
and enhancement to  the redshift space power spectrum
$P^s(k,u)$. Nevertheless, the leading order contribution to the enhancement has a  $u^4$ directional
dependence, distinctively different to the Kaiser formula. Here, $u\equiv k_z/k$, $k$ is
the amplitude of the wavevector and $k_z$
is the component along the line of sight. (III) ${\bf v}_\delta$ is of the
greatest importance for the RSD cosmology. We find that  the induced redshift clustering
shows a number of important deviations from the usual Kaiser formula. Even in the limit 
of ${\bf v}_S\rightarrow 0$ and ${\bf v}_B\rightarrow 0$, the leading order
contribution  $\propto  (1+f\tilde{W}(k)u^2)^2$. It differs from the 
Kaiser formula by a window 
function $\tilde{W}(k)$. Nonlinear evolution generically drives
$\tilde{W}(k)\leq 1$. We hence identify a significant   
systematical error causing underestimation of  the structure growth parameter
$f$ by as much as $O(10\%)$ even at relatively large scale $k=0.1h/$Mpc. (IV)
The velocity decomposition reveals the three origins of  the finger 
of God (FOG) effect and suggests to simplify and improve the modeling
of FOG by treating the three components separately. (V) We derive 
a new formula for the redshift space power spectrum. Under the velocity
decomposition scheme,   all high order Gaussian corrections and 
non-Gaussian correction of order $\delta^3$ can be taken into account {\it without
introducing extra model uncertainties}.  Here $\delta$ is the nonlinear
overdensity. (3) The velocity decomposition clarifies issues in peculiar 
velocity reconstruction through 3D galaxy distribution.  We discuss two
possible ways to carry out the 3D ${\bf 
  v}_\delta$ reconstruction. Both use the otherwise troublesome RSD in
velocity reconstruction as a valuable source of
information. Both have the advantage to render the
reconstruction of a stochastic  
3D field into the reconstruction of a deterministic window
function $W^s(k,u)$ of limited degrees of freedom.  Both can automatically and significantly alleviate the galaxy bias
problem  and, in the limit of a deterministic galaxy bias, completely overcome
it. Paper I of this series of works lays out the methodology. Companion papers
\cite{Zheng12} will extensively evaluate its performance against N-body
simulations. 
\end{abstract}
\maketitle
\section{Introduction}
The observed galaxy clustering pattern in redshift space is modified by
peculiar velocity of galaxies and shows characteristic anisotropies
\cite{Jackson72,Sargent77,Peebles80,Kaiser87,Peacock94,Ballinger96}. This
redshift space    
distortion (RSD) effect provides a promising way to measure peculiar velocity
at cosmological distance and makes itself a powerful probe of the dark
universe. It has allowed the measurement of the
structure growth rate in spectroscopic surveys such as 2dF
\cite{Peacock01,Tegmark02}, SDSS 
\cite{Tegmark04,Samushia12}, VVDS \cite{Guzzo08}, WiggleZ
\cite{Blake11b,Blake12} and BOSS 
\cite{Reid12,Tojeiro12}.  Such growth rate measurement is highly valuable in 
probing the nature of dark energy 
\cite{Amendola05,Yamamoto05,Wang08,Percival09,Song09,White09,Song10,Wang10}   
and gravity  
\cite{Zhang07c,Jain08,Linder08,Reyes10,Cai12,Gaztanaga12,Jennings12,Li12}.  In particular, with both
the expansion rate measurement from BAO
\cite{Seo03,Eisenstein05,Blake11a,Anderson12} and structure growth rate 
measurement  from RSD, spectroscopic redshift surveys are well suited to test
consistency relations in general relativity and to 
discriminate between dark energy and 
modified gravity \cite{Linder05}. For these reasons, RSD has  become 
one of the key science goals of the planned stage IV dark energy projects such
as the BigBOSS experiment \cite{BigBOSS} and the Euclid cosmology mission
\cite{Euclid}. It can also be used to measure the stochastic galaxy bias
\cite{Pen98} and the galaxy   (pairwise) velocity dispersion \cite{Jing98},
both are valuable for studying galaxy formation.  

These applications heavily relies on the RSD modeling. However, modeling
RSD to accuracy matching stage IV dark energy projects is very
challenging, especially due to three sources of nonlinear and the associated
non-Gaussianity. (1)  The nonlinear 
mapping from real space clustering to redshift space clustering. Due to this
nonlinearity, even the lowest order statistics in redshift space  involves correlations  
between density and velocity fields to infinite order
(e.g. \cite{Scoccimarro04,Seljak11, Okumura12,Okumura12b}).  (2) The nonlinear
evolution in the real space matter density and velocity fields. (3) The
nonlinear and nonlocal galaxy-matter relation.  The galaxy density bias is
known to have non-negligible nonlinearity and stochasticity
(e.g. \cite{Bonoli09}).  The galaxy velocity bias may also be needed for
precision modeling
(e.g. \cite{Desjacques10}).   To proceed, layers of 
approximations and simplifications have been made.  

Here we elaborate on some of
these approximations/simplifications with example of matter
clustering in redshift 
space. One of the most commonly used RSD  formulae connecting the isotropic real space
power spectrum $P_{\delta\delta}(k) $ to the anisotropic redshift space power spectrum
$P^s_{\delta\delta}(k,u)$  is $P^s_{\delta\delta}(k,u)\simeq
P_{\delta\delta}(k)(1+fu^2)^2D^{\rm FOG}(ku)$. It is a phenomenological
combination and extension of
the linear Kaiser effect and the finger of God (FOG) effect due to random motions. 
Here, $f\equiv d\ln D/d\ln a$ and $D\equiv D(z)$ is the linear density growth
factor at redshift $z=1/a-1$. Throughout this paper, the superscript ``s''
denotes the property in redshift space. $P^s_{\delta\delta}$ depends on both $k$ and $u\equiv
k_{\parallel}/k$. Throughout the paper we adopt the z-axis as the line of
sight, so $k_{\parallel}=k_z$.  Approximations/simplifications made include
(1)  the parallel plane  
approximation \cite{Hamilton96}, (2)  linear evolution in  the
velocity-density relation and (3) no 
stochasticity between the velocity and density field \cite{White09}.  (4) It also neglects most
high order correlations between the  density and velocity fields
\cite{Seljak11,Okumura12}.  (5) $D^{\rm FOG}(ku)$ is a phenomenological description of  the 
FOG effect, the overall damping caused by random motion. Both a
Gaussian form $D^{\rm FOG}(ku)=\exp(-(ku\sigma_v/H)^2)$ and a Lorentz form
$D^{\rm FOG}(ku)=1/(1+(ku\sigma_P/H)^2/2)$ are widely adopted. However,
the physical meaning of the velocity dispersion $\sigma_v$ and especially the pairwise
velocity dispersion $\sigma_P$ is a bit ambiguous.  Furthermore, deviations from the
above forms have been  found in
simulations \cite{Kang02}.

 Various approaches have been investigated to improve the RSD modeling. A far from complete
list includes the Eulerian
and Lagrangian perturbation  theory 
\cite{Heavens98,Scoccimarro04,Matsubara08a,Matsubara08b,Desjacques10,Taruya10,Matsubara11,Okumura11b,Sato11}, 
the halo model \cite{White01,Seljak01,Kang02,Tinker06,Tinker07}, the streaming model
\cite{Peebles80,Fisher95,Reid11}, the recently proposed
distribution function approach \cite{Seljak11,Okumura12,Okumura12b} and combinations
between them.  Furthermore, due to significant
nonlinearities involved, RSD modeling often resorts to numerical simulations
on calibration and testing 
(e.g. \cite{Kang02,Scoccimarro04,Tinker06,Okumura11a,Jennings11b,Reid11,
  Kwan12}). Despite these efforts, RSD modeling has not yet achieved the 
accuracy required for precision RSD cosmology. For example,
recent tests against N-body simulations find that the inferred $f$ can be
biased low  by $O(10\%)$ or more 
\cite{Okumura11a,Jennings11a,Jennings11b,Kwan12,Bianchi12,delaTorre12},
significantly larger than the $O(1\%)$ statistical error in $f$ for surveys
like BigBOSS and Euclid.

Since RSD is induced by peculiar velocity, it is of crucial importance to
understand the peculiar velocity field. We find that, an appropriate velocity
decomposition has the potential to simplify and improve the RSD
modeling.  We also find that the same decomposition may enable robust
reconstruction of  3D peculiar velocity
in spectroscopic surveys. It has the advantage to render the otherwise
troublesome RSD  in velocity reconstruction into valuable
source of information.  The current paper aims to lay out
the methodology of the 
proposed velocity decomposition, RSD modeling and 3D velocity
reconstruction. Extensive tests against simulations and mock catalogs are
required to clarify/justify/quantify numerous technical issues. These
numerical results will be
presented in companion papers \cite{Zheng12}. 

This paper is organized as follows. In \S \ref{sec:decomposition} we decompose
the peculiar velocity fields into three eigen-modes (${\bf v}_\delta$, ${\bf
  v}_S$ and ${\bf v}_B$) and discuss related
statistics. Among them, ${\bf v}_\delta$ is the velocity component completely
correlated with the density distribution and is the one contains most 
cosmological information. In \S \ref{sec:RD} we show that the three velocity components
affect RSD in different ways and their impacts can be treated separately. We are then
able to derive the exact RSD formula. Furthermore we propose reasonable
approximations for realistic calculation. Through this methodology, we
explicitly identify a significant systematical error in RSD
cosmology. \S \ref{sec:3D} proposes a method to 
reconstruct the 3D ${\bf v}_\delta$ from redshift surveys and the appendix 
\ref{sec:appendixD} proposes an alternative.   For brevity, the above
sections focus on the matter field. But these results can be extended to the
galaxy field straightforwardly, as briefly discussed  in \S
\ref{sec:galaxy}.   We further argue that the
velocity reconstruction is  insensitive to the galaxy bias.  At the end of
each sections, we list key statistics 
to be investigated in  future works. We also prepare four
appendices for more technical issues. 

\section{Peculiar velocity decomposition}
\label{sec:decomposition}
Any vector field can be decomposed into a irrotational (gradient) part and a rotational
(curl) part. Analogous to the electric and magnetic fields (and also the CMB
polarization field and the cosmic shear field), we denote the
first one with a subscript ``E'' and the later one with a subscript
``B''. Hence the peculiar velocity ${\bf v}$  can be decomposed as
\be
\label{eqn:EB}
{\bf v}({\bf x})={\bf v}_E({\bf x})+{\bf v}_B({\bf x})\ .
\ee
By definition, $ \nabla\times {\bf v}_E=0$ and $\nabla\cdot {\bf v}_B=0$. In
Fourier space, we have ${\bf v}_E({\bf k})=[{\bf v}({\bf k})\cdot\hat{{\bf
    k}}]\hat{{\bf k}}$ and ${\bf v}_B({\bf k})={\bf v}({\bf k})-{\bf v}_E({\bf
  k})$. 

${\bf v}_E$ can be completely described by its divergence $
\theta({\bf x})\equiv -\nabla\cdot {\bf v}({\bf x})/H\equiv -\nabla \cdot {\bf
  v}_E({\bf x})/H$.  
 To conveniently describe the velocity-density relation, we  carry out a further
decomposition,
\be
\label{eqn:mR}
 {\bf v}_E({\bf x})={\bf v}_\delta({\bf x})+{\bf v}_S({\bf x})\ \ .
\ee
Both ${\bf v}_\delta$ and ${\bf v}_S$ are irrotational ($\nabla \times {\bf v}_\delta=0$
and $\nabla \times {\bf v}_S=0$). 
 We require that  $\theta_\delta\equiv -\nabla \cdot{\bf v}_\delta/H$ is completely
correlated with the overdensity $\delta$. So we denote this component
with a subscript ``$\delta$''. In Fourier space, we then have 
\be
\label{eqn:thetam}
\theta_\delta({\bf k})=\delta({\bf k})W({\bf k})\ .
\ee
Here $W({\bf k})$ is a deterministic function of ${\bf k}$ to be determined later.

On the other hand, we require that $\theta_S\equiv
-\nabla\cdot{\bf v}_S/H$ is  uncorrelated with $\delta$. So $\langle
\theta_S({\bf x})\delta({\bf x}+{\bf r})\rangle=0$. Equivalently $\langle
\theta_S({\bf k}^{'})\delta({\bf k})\rangle=0$ for any ${\bf k}$ and ${\bf k}^{'}$. $\theta_S$ is the source
of stochasticity in the $\delta$-$\theta$ relation, so we denote this
component with a subscript ``S''.

We define the power spectrum between field $A$ and field $B$ through $\langle
A({\bf k})B({\bf k}^{'})\rangle\equiv (2\pi)^3\delta_{3D}({\bf k}+{\bf
  k}^{'})P_{AB}({\bf k})$. We often use the notation $\Delta^2_{AB}\equiv
k^3P_{AB}/(2\pi^2)$, which enters into the ensemble average $\langle A({\bf
  x})B({\bf x})\rangle=\int \Delta^2_{AB}(k)dk/k$. Through the relation
$\langle \delta({\bf k}^{'})\theta({\bf 
  k})\rangle=\langle \delta({\bf k}^{'})\theta_\delta({\bf
  k})\rangle= \langle \delta({\bf k}^{'})\delta({\bf k})\rangle W({\bf
  k})$,  we obtain
\be
\label{eqn:W}
W({\bf k})=W(k)=\frac{P_{\delta\theta}(k)}{P_{\delta\delta}(k)}\ .
\ee
Notice that, due to the isotropy of the universe, $P_{\delta\theta}({\bf
  k})=P_{\delta\theta}(k)$ and $P_{\delta\delta}({\bf
  k})=P_{\delta\delta}(k)$. So $W$ does not  depend on the direction of ${\bf
  k}$.

Hence we prove that the velocity decomposition into ${\bf v}_\delta$, ${\bf
  v}_S$ and ${\bf v}_B$ is mathematically unique,
with no assumptions on the underlying density and velocity field. But the
decomposition has strong physical motivation too.  The three components
are associated with different physical processes in the structure formation. We now
proceed to their  physical meanings.  

%%%%%%%%%%%%%%%%%%%%%%%%%%
\bfi{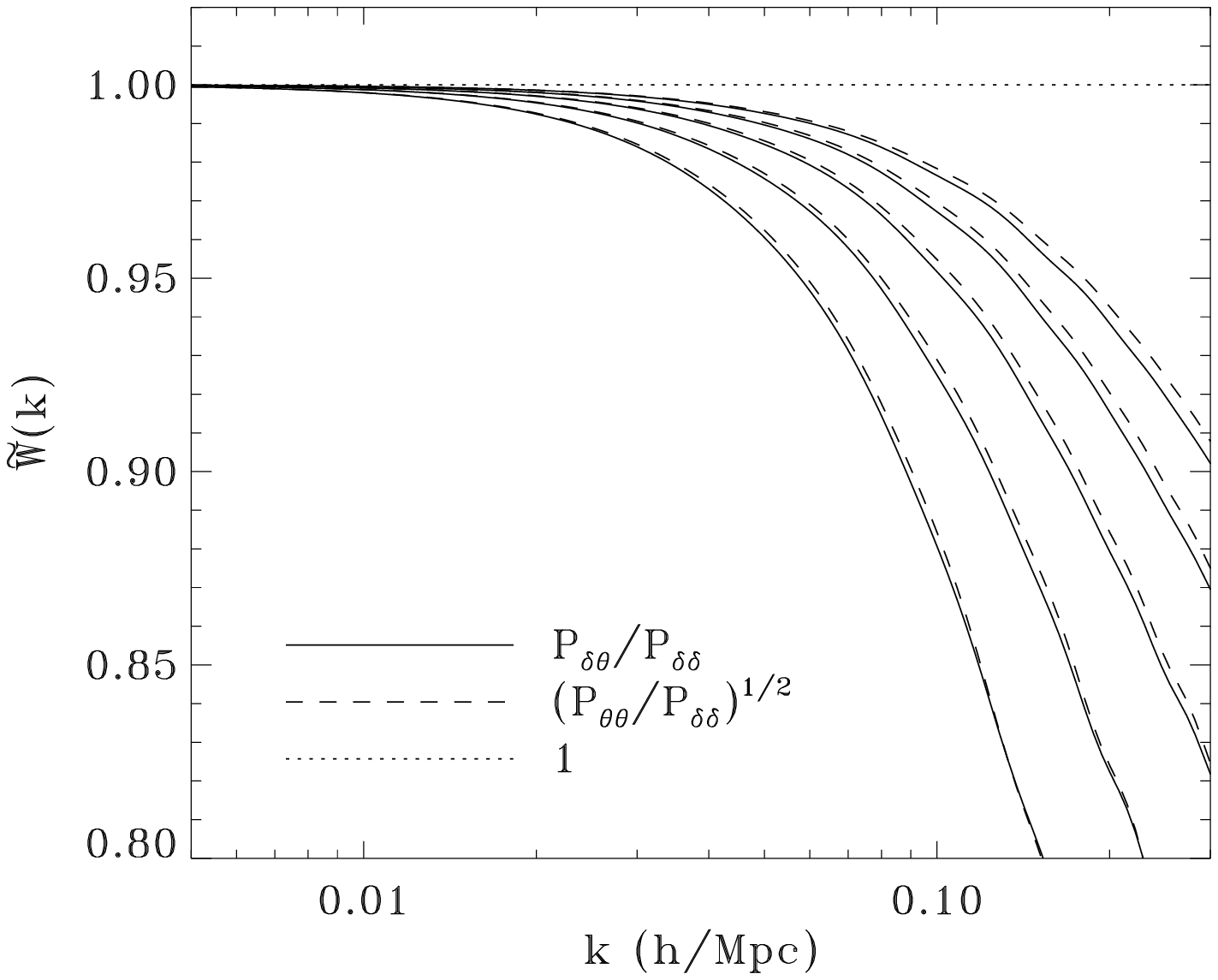}
\caption{Solid lines are $\tilde{W}(k,z)$ predicted by the third order
 Eulerian perturbation theory at redshift $z=0.0,0.5,1.0,1.5,2.0$ (bottom
  up).  We adopt a flat $\Lambda$CDM cosmology with $\Omega_m=0.26$,
  $\Omega_b=0.044$, $\Omega_{\Lambda}=0.74$, $h = 0.71$, $\sigma_8=0.8$ and $n_s=1$. 
Nonlinearity drives $\tilde{W}$ down from unity. Later we
  will show that what inferred from redshift space distortion is
  $f\tilde{W}$. So the structure growth rate $f$ can be biased low by
  $O(10\%)$, consistent with recent findings 
  (e.g. \cite{Okumura11a,Jennings11a,Jennings11b,Kwan12,Bianchi12,delaTorre12}). 
To demonstrate the ${\bf v}_S$ component, we also 
  plot $\sqrt{P_{\theta\theta}/P_{\delta\delta}}$, normalized to unity at
  $k\rightarrow 0$ (dash lines).  $\sqrt{P_{\theta\theta}/P_{\delta\delta}}=P_{\delta\theta}/P_{\delta\delta}\times
  \sqrt{1+\eta}$. $\eta\equiv
  P_{\theta_S\theta_S}/P_{\theta_\delta\theta_\delta}$ quantifies the
  relative amplitude of ${\bf v}_S$ with respect to the
  velocity component 
  ${\bf v}_\delta$. It also quantifies the velocity
  divergence-density stochasticity (Eq. \ref{eqn:r}). In the limit $k\rightarrow 0$,
  $\eta\rightarrow 0$ and $\sqrt{P_{\theta\theta}/P_{\delta\delta}}\rightarrow
  P_{\delta\theta}/P_{\delta\delta}$. The nonlinear evolution generates ${\bf
    v}_S$ and causes the two sets of curves to deviate from each other.  Since
  the third order Eulerian
  perturbation theory has limited range of applicability, numerical results
  shown in this plot are mainly presented to demonstrate the major impacts of
  the nonlinear evolution such as driving $\tilde{W}$ down from unity and
  driving $\eta$ up from 
  zero. Robust quantifications of $\tilde{W}$ and $\eta$ will be presented in
  companion papers \cite{Zheng12}.  \label{fig:W}} 
\efi
%%%%%%%%%%%%%%%%%%%%%%%%%%

\subsection{The ${\bf v}_\delta$ field}
\label{subsec:m}
In the limit $k\ll k_{\rm NL}$, ${\bf v}_\delta$ is the only velocity component.  Here
$k_{\rm NL}$ is the nonlinear scale, defined through
$\Delta^2_{\delta\delta}(k_{\rm NL})=1$. ${\bf v}_B$ is a
decay mode before shell crossing  so it is negligible
in the linear regime \cite{Bernardeau02,Pueblas09}. On the other hand,  in the limit
$k\ll k_{\rm NL}$, ${\bf v}_E$ is 
completely correlated with the density  
field, with a deterministic relation $\theta=f\delta$. So ${\bf
  v}_S$ vanishes too. ${\bf v}_\delta$, being  the most linear velocity
component, is of the greatest
interest to cosmology. 

Nevertheless, nonlinear evolution leaves non-negligible imprints in the ${\bf
  v}_\delta$ field.   A crucial point is that nonlinear 
evolution affects the density field and the velocity field in different or
even opposite ways  (e.g. \cite{Bernardeau02,Scoccimarro04}). The third
order Eulerian perturbation shows that, when the 
effective power index $n_{\rm eff}\agt -1.9$, nonlinear evolution actually
suppresses the velocity growth, while enhances the overdensity growth
(Fig. 12, \cite{Bernardeau02}). Even for  $n_{\rm eff}\alt -1.9$, $\theta$
grows more slowly than the overdensity  (Fig. 12,
\cite{Bernardeau02}). Furthermore, nonlinear evolution generates a stochastic
velocity component ${\bf v}_S$, further reducing $\theta_\delta$ with respect
to $\delta$. In the deeply nonlinear regime, after many orbit crossings, the
velocity field eventually loses  its correlation with the density field and we
expect ${\bf v}_\delta\rightarrow 0$.

Hence  we should use Eq. \ref{eqn:W} to describe the $\theta_\delta$-$\delta$
relation, instead of the linear relation $\theta_\delta=f\delta$. For the
convenience of highlighting the deviation 
from the linear relation, we  define the normalized $\tilde{W}$
through
\be
\label{eqn:tildeW}
\tilde{W}(k)\equiv \frac{W(k)}{W(k\rightarrow 0)}=\frac{W(k)}{f}=\frac{1}{f}\frac{P_{\delta\theta}(k)}{P_{\delta\delta}(k)}\ .
\ee

$\tilde{W}$ calculated using the third order Eulerian perturbation theory is shown in
Fig. \ref{fig:W}. In companion papers \cite{Zheng12} we will numerically
evaluate this key quantity using high resolution N-body simulations. In the limit $k\rightarrow 0$, $\tilde{W}\rightarrow 1$ as
expected. But nonlinear evolution soon drives $\tilde{W}$ to deviate
$\tilde{W}<1$. Even at relatively high redshift $z=2$ and pretty linear scale
$k=0.1h/$Mpc, the deviation already reaches $O(1\%)$.  The derivation
increases towards low redshift and exceeds $10\%$ at $z\alt 0.5$. 

 To better understand this
behavior, we can express $\tilde{W}$ 
as 
\be
\tilde{W}(k)=r_{\delta
  \theta}(k)\sqrt{\frac{P_{\theta\theta}(k)}{f^2P_{\delta\delta}(k)}}\leq
\sqrt{\frac{P_{\theta\theta}(k)}{f^2P_{\delta\delta}(k)}}\leq 1\ .
\ee
Here, $r_{\delta \theta}(k)$ is the cross correlation coefficient
between $\delta$ and $\theta$. By definition, $r_{\delta\theta}(k)\leq 1$.
Given the fact that velocity growths more slowly the density,
$P_{\theta\theta}<f^2P_{\delta\delta}$. So we expect $\tilde{W}<1$ in
nonlinear regime.  Furthermore,  when $k\gg k_{\rm NL}$,
$r_{\delta\theta}\rightarrow 0$. So we expect $\tilde{W}\rightarrow 0$ when
$k\gg k_{\rm NL}$. 

The recognition of the velocity component ${\bf v}_\delta$  has several
important applications. (1) It lays out a promising way to reconstruct the
3D peculiar velocity from 3D density distribution, which is accessible from
galaxy spectroscopic redshift surveys. $\tilde{W}$ mimics a window function
with a smoothing scale comparable to the nonlinear scale. It exerts on the
density field and suppresses small scale inhomogeneities so that the smoothed
density field provides a honest estimation on the underlying velocity field
${\bf v}_\delta$.  One can find a similar window function exerting on the
redshift space density, which is directly observable.  Hence the
reconstruction of the stochastic 3D vector field is simplified to the
reconstruction of a deterministic window function with limited degrees of
freedom.  This is one of the most
important applications of the velocity decomposition proposed in this paper.
Later in \S \ref{sec:3D} we will present more detailed investigation.  (2) It simplifies the RSD modeling. Since $\tilde{W}\rightarrow 0$ toward
  small scales, nonlinearity and non-Gaussianity in the ${\bf v}_\delta$ field
  are significantly suppressed.  (3) It identifies a severe systematical error
  in RSD cosmology.  Later in \S \ref{subsec:RDvdelta} we will show that RSD is
  determined by $f\tilde{W}$ instead of $f$. If the factor $\tilde{W}$ is not
  included in the RSD cosmology, $f$ will be biased low by  $\sim 2\%$ at
  $z=2$ and $\sim 10\%$ at $z=0$,   even if we restrict the analysis to $k\alt
  0.1h/$Mpc. To our best  knowledge, this is the first time such systematical 
error is diagnosized explicitly. This source of systematical error could explain recent findings
  of $O(10\%)$ underestimation in $f$ inferred from simulated RSD
  data \cite{Okumura11a,Jennings11a,Jennings11b,Kwan12,Bianchi12,delaTorre12}.
  We will present more discussions on its cosmological implications in \S \ref{subsec:cosmology}.

\subsection{The ${\bf v}_S$ field}
\label{subsec:S}
To the opposite of ${\bf v}_\delta$, ${\bf v}_S$
vanishes in the linear regime and begins to grow due to the nonlinear evolution.
For this reason, it lacks large scale  power and hence its correlation length
is smaller than that of ${\bf  v}_\delta$. For the same
reason, it is intrinsically non-Gaussian. 
Being uncorrelated to the density field, ${\bf v}_S$  induce stochasticities
in the velocity-density 
relation. We have the relation
\be
\label{eqn:r}
r_{\delta\theta}(k)\equiv
\frac{P_{\delta\theta}(k)}{\sqrt{P_{\delta\delta}(k)P_{\theta\theta}(k)}}=\frac{1}{\sqrt{1+\eta(k)}}\ ,
\ee
where $\eta(k)\equiv P_{\theta_S\theta_S}(k)/P_{\theta_\delta\theta_\delta}(k)$. When
${\bf v}_S$ overwhelms over ${\bf v}_\delta$ ($\eta\gg 1$),
$r_{\delta\theta}\rightarrow 0$ and the velocity field loses its correlation
with the density field. 

Based on the  third order perturbation calculation, we find that
$P_{\theta_S\theta_S}$ already reaches $\simeq 1\%$ of
$P_{\theta_\delta\theta_\delta}$ (namely $\eta\simeq 1\%$) at $k=0.1h/$ Mpc
and  $z\alt 1$, as can be  inferred from Fig. \ref{fig:W}.  As expected,  the 
situation is less severe at higher redshifts. But even at $z=2$, $\eta\simeq
1\%$ at $k=0.2h/$Mpc.  Hence in general, ${\bf v}_S$ is a
non-negligible velocity component even at scales which are often considered as
linear. Nevertheless, it
is still subdominant to ${\bf v}_\delta$ at scale $k\alt 1 h/$Mpc, as one can
infer from Fig. 1 of \cite{White09}, in combination with our Eq. \ref{eqn:r}. Our
numerical evalulations obtain similiar results  \cite{Zheng12}. 

What cosmological information does ${\bf v}_S$ contain? One particularly interesting
piece of information is likely  the nature of gravity. For modified
gravity models to pass the solar system tests and to drive the late
time cosmic acceleration, gravity must behave upon the  environment (e.g. review articles 
\cite{Jain10,Clifton12}). Such environmental
dependence often becomes prominent in the nonlinear regime, so it  can
significantly affects the velocity divergence
(e.g. \cite{Li12}). Arisen from nonlinear evolution, ${\bf  v}_S$
would be sensitive to this generic feature of modified gravity, making it attractive for testing
gravity. 
 
In \S \ref{sec:RD} we will show that  the ${\bf v}_S$ induced RSD differs
from the ${\bf v}_\delta$ induced RSD (\S \ref{sec:RD}). We argue that
these differences can be used to separate the different velocity components
and to make {\it statistical} measurement of ${\bf v}_S$ possible.  

\subsection{The ${\bf v}_B$ field}
\label{subsec:B}

${\bf v}_B$ decays as long as the single fluid approximation for the
dark matter distribution holds \cite{Bernardeau02}. ${\bf v}_B$ grows only
when the nonlinearity is sufficiently large that shell 
crossing happens. In the
deeply nonlinear regime,  we may even expect equi-partition in the velocity
distribution and ${\bf v}_B$ can dominate over ${\bf v}_E$. So we expect that its power
concentrates at even smaller scales than ${\bf v}_S$.    Indeed, numerically
studies \cite{Pueblas09} show 
that the power of this velocity component concentrates at small scales, with
r.m.s. much smaller than that of ${\bf v}_E$.  Our numerical evalulations
found similar behavior  \cite{Zheng12}.  Later in \S \ref{subsec:RDvB}
we will show that these behaviors significantly simplify the modeling of ${\bf
  v}_B$ induced RSD. We refer the readers
to \cite{Pueblas09} and references therein for in depth study of ${\bf
  v}_B$. In \cite{Zheng12}, we will present our numerical evaluations on RSD
related statistics of ${\bf v}_B$.

\subsection{Statistical description of the three velocity components}
\label{subsec:velocitystatistics}
This paper is  not at a position to calculate the statistics of these velocity
components. Instead, we present some general discussions here and postpone any
quantitative calculations into future works. First, since the three velocity
components have different origins, the halo model scenario
\cite{Jing98,Cooray02,Yang03,Zheng05} provides an unified way to describe the 
three components. Nevertheless, a number of extensions/corrections may be
necessary. (1)  Since the linear perturbation theory fails  to describe the
emergence of ${\bf v}_S$ and deviations from the linear  theory prediction of
the halo bulk motion have been diagnosized 
\cite{Hamana03}, a natural extension to describe the halo bulk motion  is the third order
perturbation. (2) Peculiar velocity has been treated as the sum of the
halo bulk velocity and the random velocity   inside of the virialized halos
\cite{Sheth01a}. It may be extended to include 
the more complicated motion around the halo outskirt, which may be a
significant source of ${\bf v}_B$ \cite{Pueblas09}.  Nevertheless, treatment of ${\bf v}_B$ is trickier and we refer the readers to
\cite{Pueblas09} for detailed discussions.

One important statistics relevant to RSD is the velocity correlation
function. Due to symmetry considerations, it can be
decomposed as  \cite{Peebles80}
\be
\langle v_i({\bf x}_1)v_j({\bf
  x}_2)\rangle=\psi_{\perp}(r)\delta_{ij}+\left[\psi_{\parallel}(r)-\psi_{\perp}(r)\right]\frac{r_ir_j}{r^2}
\ . 
\ee
Here ${\bf r}\equiv {\bf x}_1-{\bf x}_2$ is the pair seperation vector and
$i=x,y,z$ is the Cartesian axes.  $\psi_{\parallel}$ is the
correlation function when both $v_i$ and $v_j$ are  along ${\bf
  r}$. $\psi_{\perp}$ is the one when both 
velocities are perpendicular to  ${\bf r}$.

$\psi_{\parallel}$ and $\psi_{\perp}$ are not
independent. For a potential flow (${\bf
  v}_\delta$ and ${\bf v}_S$), we have the textbook result  \cite{Peebles80}
\ba
\psi_{\parallel}(r)&=&\frac{d(r\psi_{\perp}(r))}{dr}\ ,\\
\psi_{\perp}(r)&=&H^2\int \Delta^2_{\theta\theta}\frac{\sin(kr)}{kr}\frac{dk}{k^3}\ .
\ea
Here we have defined the velocity power spectrum through $\langle {\bf v}({\bf
  k})\cdot {\bf v}({\bf k}^{'})\rangle\equiv (2\pi)^3\delta_{3D}({\bf k}+{\bf
  k}^{'})P_{vv}(k)$. Its covariance is defined as $\Delta^2_{vv}\equiv
P_{vv}k^3/2\pi^2$.

On the other hand, ${\bf v}_B$ does not follow this relation. Using the fact that ${\bf
  v}_B$ can be expressed as the vorticity of a vector ${\bf A}$ (${\bf
  v}_B=\nabla \times {\bf A}$), we derive the
following relations,
\ba
\psi_{\perp}(r)&=&\psi_{\parallel}(r)+\frac{1}{2}r\frac{d\psi_{\parallel}(r)}{dr}\ ,\\
\psi_{\parallel}(r)&=&\int
\Delta^2_{v_Bv_B}(k)\frac{dk}{k}\frac{1}{3}\left[\frac{\sin
    kr}{(kr)^3}-\frac{\cos kr}{(kr)^2}\right]\ .
\ea

\subsection{To do list}
In companion papers \cite{Zheng12}, we will use N-body simulations to
numerically evaluate statistics of these velocity fields. An incomplete list includes
\bi
\item The power spectra $P_{v_\delta v_\delta}(k,z)$, $P_{v_S v_S}(k,z)$ and
  $P_{v_Bv_B}(k,z)$. Their scale and redshift dependences are helpful to
  understand the impact of nonlinear evolution. 
\item The correlation function $\psi_{v_\delta v_\delta}(r,z)$, $\psi_{v_S v_S}(r,z)$ and
  $\psi_{v_B v_B}(r,z)$, of the ``$\parallel$'' mode and the ``$\perp$'' mode. These statistics quantify the correlation length of the
  three velocity components and tell us at which separation we can treat the
  velocities at two positions as independent. 
\item The PDFs of ${\bf v}_\delta$, ${\bf v}_S$ and
  ${\bf v}_B$.  The PDFs determine the overall damping (FOG) to the redshift power
  spectrum.  We will also calculate the
  cumulants to quantify non-Gaussianity of these velocity components.
\item $\tilde{W}(k,z)$. As we have addressed, $\tilde{W}$ is of crucial importance
  in the 3D velocity reconstruction and in the RSD cosmology. Through N-body
  simulations, we will robustly measure its dependence on $k$ and
  $z$. Furthermore, we want to quantify 
  the $\tilde{W}(k,z)$-$\Delta^2_{\delta\delta}(k,z)$ relation to better
  understand its dependence on the nonlinearity. Especially, we
  want to know if it can be well
  approximated by a simple function with only a few parameters. 
\ei

\section{Modeling the redshift space distortion}
\label{sec:RD}
Now we proceed to the RSD modeling with the aid of the proposed velocity
decomposion.  Following \cite{Scoccimarro04}, we utilize the matter
conservation ($(1+\delta^s({\bf  
  x}^s))d^3{\bf x}^s=(1+\delta({\bf x}))d^3{\bf x}$) to derive
\be
\delta^s({\bf k})=\int \left[1+\delta({\bf x})\right]\exp\left(i\frac{k_zv_z}{H}\right)\exp(i{\bf k}\cdot{\bf
  x})d^3{\bf x}
\ee
In the above expression, we have neglected a Dirac funtion which only shows up
when ${\bf k}=0$,  irrelevant to our calculation. 
This equation adopts the plane parallel approximation and adopts
the line of sight as the z-axis. It also assumes no multiple
streaming. Extention to this more complicated situation can be done by  the
phase space distribution 
function approach \cite{Seljak11,Okumura12,Okumura12b}. Phenomenologically
speaking, multiple streaming can be described by an overall damping to the
redshift space power spectrum. Hence we do not expect major
changes to the results presented in this paper. 

The redshift space power spectrum is given by \cite{Scoccimarro04}
\ba
\label{eqn:ps0}
P_{\delta\delta}^s({\bf k})=\int \left\langle (1+\delta_1)(1+\delta_2)\exp\lambda\right\rangle  \exp(-i{\bf k}\cdot{\bf r})d^3{\bf r}
\ea
where $\delta_i\equiv \delta({\bf x}_i)$ and ${\bf x}_2\equiv {\bf x}_1+{\bf
  r}$. The ensemble average $\langle \cdots\rangle$ is averaged over all ${\bf
  x}_1$ with fixed ${\bf r}$. For brevity, we have denoted $\lambda\equiv
ik_z(v_{1z}-v_{2z})/H$. We also denote $\lambda_{\alpha} \equiv
ik_z(v_{1z,\alpha}-v_{2z,\alpha})/H$ where $\alpha=\delta,S,B$. Due to the
axial symmetry along the line of sight, $P_{\delta\delta}^s({\bf k})$ only depends on
$k$ and $u$. Hereafter we often write it as $P_{\delta\delta}^s(k,u)$.

An immediate result is that $\delta^s({\bf
  k}_{\perp},k_z=0)=\delta({\bf k}_{\perp},k_z=0)$ and $P^s_{\delta\delta}({\bf
  k}_{\perp},k_z=0)=P_{\delta\delta}(k_{\perp})$. A similar relation holds for
the bispectrum, $B^s_3({\bf 
  k}_1,{\bf k}_2,{\bf k}_3)=B_3({\bf
  k}_1,{\bf k}_2,{\bf k}_3)$ when $k_{1,z}=k_{2,z}=k_{3,z}=0$. Later we will
show that these relations are very useful in RSD modeling. Notice that no such 
relations exist in real space. For example, $\xi^s({\bf r}_{\perp},r_z=0)\neq
\xi(r_{\perp})$. 

\subsection{The role of ${\bf v}_B$}
\label{subsec:RDvB}
Now we will apply the velocity decomposition to Eq. \ref{eqn:ps0}. 
Since only ${\bf v}_\delta$ is correlated with the overdensity,
Eq. \ref{eqn:ps0} can be reduced to 
\ba
\label{eqn:ps1}
P^s_{\delta\delta}(k,u)&=&\int \left\{\left\langle
(1+\delta_1)(1+\delta_2)\exp\lambda_\delta\right\rangle \right.\\
&\times&\left. D_S(k_z,{\bf r})D_B(k_z,{\bf r}) \right\}
\exp(-i{\bf k}\cdot{\bf r})d^3{\bf r}\ .\no 
\ea
The two function $D_S(k_z,{\bf r})$ and $D_{B}(k_z,{\bf
  r})$ completely describe the redshift  distortion caused by ${\bf v}_S$
and ${\bf v}_B$ respectively. 
\be
D_S(k_z,{\bf r})\equiv\left\langle
\exp\left(i\frac{k_z(v_{1z,R}-v_{2z,R})}{H}\right)\right\rangle\ ,
\ee
\be
D_B(k_z,{\bf r})\equiv\left\langle
\exp\left(i\frac{k_z(v_{1z,B}-v_{2z,B})}{H}\right)\right\rangle\ .
\ee
Due to the symmetry between $k_z$ and $-k_z$, $D_{S,B}(k_z,{\bf
  r})=D_{S,B}(|k_z|,{\bf r})$.  

We have $D_S\leq 1$ and $D_B\leq 1$, so both ${\bf v}_S$
and ${\bf v}_B$ suppress the clustering presented by terms inside of the
bracket of Eq. \ref{eqn:ps1}. However, ${\bf v}_S$ and ${\bf v}_B$ can have
their own clusterings and hence can in principle increase the overall
clustering in redshift space.  We can also define a corresponding function for ${\bf
  v}_\delta$, 
\be
D_\delta(k_z,{\bf r})\equiv\left\langle
\exp\left(i\frac{k_z(v_{1z,\delta}-v_{2z,\delta})}{H}\right)\right\rangle\ .
\ee
But since ${\bf v}_\delta$ is correlated with the density field, its role in
RSD is too complicated to be described by a single function $D_\delta$. 

 In general,  these functions depend on the pair separation ${\bf r}$, due to
 correlations in the corresponding velocity fields. To describe this effect,
 we define 
\be
1+\epsilon_{\alpha}({\bf r},k_z)\equiv \frac{D_{\alpha}(k_z,{\bf r})}{D_{\alpha}(k_z,r\rightarrow \infty)}\ .
\ee
Here,   $\alpha=\delta$, $S$, $B$ respectively. The limit $r\rightarrow
\infty$ corresponds to the limit of no velocity 
correlation.  In this limit, peculiar velocity only causes damping (FOG). Hence
we denote $D$ at this limit with a superscript ``FOG'', 
\ba
D_{\alpha}(k_z,r\rightarrow \infty)&\equiv& D^{\rm FOG}_{\alpha}(k_z)\\
&=&
\left|\left\langle \exp\left(i\frac{k_zv_{z,\alpha}}{H}\right)
\right\rangle\right|^2\ . \no
\ea

As discussed in \S \ref{sec:decomposition}, ${\bf v}_B$ arises mostly in the
 deeply nonlinear regime, so we expect a correlation length shorter than
 scales of interest for RSD cosmology. Hence we can neglect correlation in the
 ${\bf v}_B$ field. Namely, we can make the
 approximation
\ba
\label{eqn:DB}
\epsilon_B\simeq 0\ \ ,\ \ D_{B}(k_z,{\bf r})\simeq D^{\rm
  FOG}_{B}(k_z)\ .
\ea
Since $D^{\rm FOG}_{B}$ is independent of ${\bf r}$,  Eq. \ref{eqn:ps1}
is simplified as
\ba
\label{eqn:ps2}
P^s_{\delta\delta} (k,u)&=&\left\{\int \left\langle
(1+\delta_1)(1+\delta_2)\exp\lambda_\delta\right\rangle\right. \\
&&\left.  D_S(k_z,{\bf r}) \exp(-i{\bf k}\cdot{\bf r})d^3{\bf r}\right\}D^{\rm FOG}_B(k_z)\ . \no 
\ea
The damping function $D^{\rm FOG}_B$ is related to its velocity PDF through
\ba
\sqrt{D^{\rm FOG}_B(k_z)}&=&\int_{-\infty}^{\infty} \exp\left(i\frac{k_zv_{z,B}}{H}\right)
P_B(v_{z,B})dv_{z,B}\no \\
&=&\int _{-\infty}^{\infty}\cos\left(\frac{k_zv_{z,B}}{H}\right)
P_B(v_{z,B})dv_{z,B} \ . \no
\ea
Through the cumulant expansion theorem, we can express $D^{\rm FOG}_B$ in
cumulants of $v_{z,B}$. The cumulant expansion
theorem states that
\ba
\label{eqn:cumulantexpansion}
\left\langle
\exp\left(i\frac{k_zv}{H}\right)\right\rangle=\exp\left(\sum_{n=1}^{\infty}
\left(\frac{ik_z}{H}\right)^n\frac{\langle v^n\rangle_c}{n!}\right)\\
=\exp\left(-\frac{k_z^2\sigma_v^2}{2H^2}\right)\exp\left(\sum_{j\geq
    2}^{\infty} (-1)^j\left(\frac{k_z\sigma_v}{H}\right)^{2j}
\frac{K_{2j}}{(2j)!}\right) \no \\
=\exp\left(-\frac{x}{2}\left[1-\frac{K_4}{12}x+\frac{K_6}{360}x^2+\cdots \right]\right) \no \ .
\ea
Here, $v$ denotes $v_{z,B}$.  $\langle \cdots\rangle_c$ is
the corresponding cumulant of the quantity inside of the bracket.  For a Gaussian 
velocity distribution, we have $\langle v^j\rangle_c=0$ for $j\geq 3$ and
recover the Gaussian FOG.  For non-Gaussian distribution, higher order terms
show up. Since 
$\langle v^{2j+1}\rangle_c=0$, only even cumulants contribute. $K_j\equiv
\langle v^j\rangle_c/\sigma_v^j$ is the reduced cumulant and $x\equiv 
(k_z\sigma_v/H)^2$.   

Non-Gaussianity in ${\bf v}_B$ could be significant and we may expect that
high order cumulants must be included to robustly model $D^{\rm
  FOG}_B$. Fortunately, the reality can be much simpler due to the fact that
${\bf v}_B$ has a $\sigma_v$much smaller than that of ${\bf v}_E$
\cite{Pueblas09,Zheng12}.  Hence for the
scales of interest (e.g. $k<1h/$Mpc), we have $x\ll 1$. Then the ${\bf
  v}_B$ induced FOG can be well described by the
following Gaussian form,  
\be
\label{eqn:FOGB}
D_B^{\rm
  FOG}(k_z)\simeq \exp \left(-\frac{k_z^2\sigma_{v_B}^2}{H^2}\right)\ .
\ee
$\sigma_{v_B}$ is hard to calculate from first principle, so it shall be
treated as a free parameter to be fitted by the data. 

\subsection{The role of ${\bf v}_S$}
\label{subsec:RDvS}
The situation for ${\bf v}_S$ is more complicated. Since a significant
fraction of ${\bf v}_S$ comes from bulk motion, ${\bf v}_S$ can be
still correlated over $O(10)$ Mpc separation. So we are no longer able to
make the approximation $\epsilon_S=0$.  Instead, we have, with the aid of the cumulant
expansion theorem,   
\ba
\label{eqn:DR}
1+\epsilon_S({\bf r},k_z)=\exp\left[\frac{k_z^2\langle
    v_{1z,S}v_{2z,S}\rangle}{H^2}\right]\\
\times e^{[k_z^4\langle
    3v^2_{1z,S}v^2_{2z,S}-2(v^3_{1z,S}v_{2z,S}+1\leftrightarrow
    2)\rangle_c/12H^4+\cdots]}\no \\ 
=1+\frac{k_z^2\langle
    v_{1z,S}v_{2z,S}\rangle}{H^2}+O(v_S^4)\ . \no
\ea
Due to the symmetry $v\leftrightarrow -v$, odd terms in $v$ do not show up in
the above equation. Plugging it into Eq. \ref{eqn:ps2}, we find that the  leading order
contribution to $P^s_{\delta\delta}$ is $P_{\theta_S\theta_S}u^4$
(Eq. \ref{eqn:RD2}), which enhances 
$P^s_{\delta\delta}$. Notice that, since ${\bf v}_S$ is uncorrelated with the
density field, it does 
not contribute a $u^2$ term as ${\bf v}_\delta$ does.  Higher order
contributions come from both fourth and higher order velocity correlations
(Eq. \ref{eqn:DR}) and from the convolution of $\epsilon_S$ with $\langle
(1+\delta_1)(1+\delta_2)\exp\lambda_\delta\rangle$ (Eq. \ref{eqn:ps1}). These
corrections are of the order $\delta^4$ or higher. We group all these
contributions into a single function $C_S(k,u)$ (Eq. \ref{eqn:RD2}). Later we argue that we may be 
able to  set $C_S=0$ in the RSD modeling. 

On the other hand, ${\bf v}_S$ causes an overall damping, characterized by
$D^{\rm FOG}_S$, 
\ba
\sqrt{D^{\rm FOG}_S(ku)}&=&\int_{-\infty}^{\infty} \exp\left(i\frac{kuv_{z,S}}{H}\right)
P_S(v_{z,S})dv_{z,S}\no \\
&=&\int_{-\infty}^{\infty} \cos\left(\frac{kuv_{z,S}}{H}\right)
P_S(v_{z,S})dv_{z,S}\ . \no
\ea
Since ${\bf 
  v}_S$ arises from nonlinear evolution, it is intrinsically non-Gaussian. We
will  check if $D^{\rm FOG}$ can be well described by the first  
several cumulants, with the aid of the cumulant expansion theorem
(Eq. \ref{eqn:cumulantexpansion}).  

\subsection{The role of ${\bf v}_\delta$}
\label{subsec:RDvdelta}
Due to coupling between the velocity and the density fields, redshift distortion induced by
${\bf v}_\delta$ is complicated, although analytical expression in the Gaussian
limit exists \cite{Scoccimarro04}.  We group the Gaussian terms with order
higher than $\delta^2$ as $C_G$. We also derive the
non-Gaussian corrections and group them as $C_{NG}$ in the appendix. Both
terms are consequeces of the ${\bf 
  v}_\delta$-$\delta$ correlation. 

In Fourier space, we have  
\ba
\label{eqn:RD2}
P^s_{\delta\delta}(k,u)&=&\left\{P_{\delta\delta}(k)(1+f\tilde{W}(k)u^2)^2+u^4P_{\theta_S\theta_S}(k)
\right. \no\\
&&\left. +C_{NG}(k,u)+C_{G}(k,u)+C_S(k,u)\right\}\no \\
&&\times D^{\rm FOG}_\delta(ku)D^{\rm
  FOG}_S(ku)D^{\rm FOG}_B(ku) \ .
\ea
Here, $C_{G}$ and $C_{NG}$ are the Fourier transforms of the corresponding
terms in real space ($C_{G}({\bf r},k_z)$, Eq. \ref{eqn:CGreal} and $C_{NG}({\bf r},k_z)$,
Eq. \ref{eqn:CNGreal}).

${\bf v}_\delta$ also causes a FOG effect, described by $D^{\rm FOG}_\delta$. ${\bf v}_\delta$ is largely Gaussian due to the suppression of $\tilde{W}<1$
in the nonlinear and non-Gaussian regime. However, due to the large amplitude
of $\sigma_{v_\delta}$,  it is unclear whether we can neglect the $K_4$ and/or
$K_6$ corrections. Here $\sigma_{v_\delta}$ is the  the one
dimension velocity dispersion of ${\bf v}_\delta$, 
\ba
\label{eqn:sigmav}
\sigma^2_{v_\delta}&\equiv& \langle v_{\delta,z}^2\rangle=\frac{1}{3}\int
\Delta^2_{v_\delta v_\delta}(k)\frac{dk}{k}\\
&=&\frac{1}{3}\int \frac{H^2}{k^2}
\Delta^2_{\delta\delta}(k)W^2(k)\frac{dk}{k} \ . \no
\ea

The $C_{NG}$, $C_G$ and $C_S$ terms are the sums of infinite
power series of $\delta$. To carry out realistic calculation, we need to
truncate them in a reasonable way.  The first line terms in the right hand side of Eq. \ref{eqn:RD2} exhaust
contributions of the order $\delta^2$. It can be rewritten in a more familiar
form $P_{\delta\delta}+2u^2P_{\delta\theta}+u^4P_{\theta\theta}$
(e.g. \cite{Scoccimarro04}) and confirmes previous results. The second  line
terms are higher   
order in $\delta$. The leading order term of $C_{NG}$ is $\propto \delta^3$, while
those of $C_{G}$ and $C_S$ are $\propto \delta^4$. In this sense, for
Eq. \ref{eqn:RD2} to be complete at the order of $\delta^3$,  we can set
$C_G=0$ and $C_S=0$. 

But in term of the linear density $\delta_L$, the
situation is different.  (1) $P_{\theta_S\theta_S}$
vanishes in the linear perturbation theory, but emerges in the second or higher
order Eulerian perturbation 
theory. So $P_{\theta_S\theta_S}$ itself is of the order $\delta_L^4$ and $C_S$ is of the order $\delta_L^6$. (2) In
the Eulerian  perturbation theory, leading order terms of $C_G$ and 
$C_{NG}$ are of the order  $\delta_L^4$.  So up to the order $\delta^4_L$,
we can set $C_S=0$. 

Hence no matter in power series of $\delta$ or in power series of $\delta_L$,
we can set $C_S=0$ when calculating the leading order corrections to the usual
RSD formula. But whether or not we shall set $C_G=0$ is an issue for numerical
examination.

The inclusion of $C_G$ and $C_{NG}$ may appear to introduce more difficulties
and uncertainties to the RSD modeling. However, this is not the case.  We will
show that,  {\it the inclusion of $C_G$ and  leading order
terms in $C_{NG}$ does not
introduce extra degrees of freedom in the RSD modeling}. 
So the inclusion of these terms is capable of  reducing systematical
errors, without degrading cosmological parameter constraints. This is
definitely a desirable property, made possible by the proposed velocity
decomposition.  

\subsubsection{The $C_{G}$ correction}
\label{subsubsec:CG}
$C_G$ is an analytical (but nonlinear) function of the two point density and velocity (${\bf
  v}_\delta$) correlation functions (Eq. \ref{eqn:CGreal}). Since
$P_{\delta\delta}$ is directly measurable from the $k_z=0$ Fourier modes in
redshift surveys, $C_G$ up to any order can be calculated strictly without
introduce any extra  parameters. 

$C_G$ does not contain terms of odd order in $\delta$. So we can split $C_G$ as
$C_G(k,u)=\sum_j C_{G,2j}(k,u)$ where $j=2,3\cdots$.  In reality, we may only need the
leading order 
\ba
C_{G,4}(k,u)&=&\int \frac{d^3{\bf
    k}_1}{(2\pi)^3}\ P_{\delta\delta}(k_1)P_{\delta\delta}(k_2)\frac{k_{1z}k_{2z}k^2_zW_1W_2}{k_{1}^2k_{2}^2}\no
\\
&\times& G({\bf k},{\bf k}_1,{\bf k}_2)\ .
\ea
Here, ${\bf k}_2={\bf k}+{\bf k}_1$ and $W_i\equiv W(k_i)$. The kernel $G$ is 
\be
G=\frac{k_{1z}W_1/k_1^2}{k_{2z}W_2/k_2^2}-1+\frac{2k_{1z}k_zW_1}{k_1^2}+\frac{k_{1z}k_{2z}k^2_zW_1W_2}{2k_{1}^2k_{2}^2}\ .
\ee
With $P_{\delta\delta}$ an observable and $W$ a function to be fitted anyway,
calculating $C_G$ requires no extra free parameters and hence does not induce
new model uncertainties. 

 \subsubsection{The $C_{NG}$ correction}
\label{subsubsec:CNG}
$C_{NG}(k,u)=\sum_{j\geq 3}
C_{NG,j}(k,u)$ contains connected part of $j$-th order  correlation
($\langle v_\delta^j\rangle$, $\langle \delta v_\delta^{j-1}\rangle$ and $\langle \delta\delta
v_\delta^{j-2}\rangle$) to $j\rightarrow \infty$ (Eq. \ref{eqn:CNGreal}). These terms are the
consequence of nonlinear 
and non-Gaussian  evolution, so we expect them to become non-negligible only
in nonlinear regime and may become dominant in deeply nonlinear regime. However, since
$\tilde{W}\rightarrow 0$ in deeply nonlinear regime, their contributions to
$P_{\delta\delta}^s$ are significantly suppressed.  For the same reason, $C_{NG,j+1}$ is
suppressed by a factor $\sim \tilde{W}$, with respect to $C_{NG,j}$. Hence we
propose to keep only the $j=3$ term and neglect higher order corrections.

Fourier transforming $C_{NG,3}({\bf r},k_z)$ (Eq. \ref{eqn:CNGreal}), we
obtain
\ba
\label{eqn:C3Fourier}
C_{NG,3}(k,u)&=&\int \frac{d^3{\bf
    k}_1}{(2\pi)^3} B_3({\bf k}_1,{\bf k}_2,{\bf
  k}) \frac{k_{1z}k_z}{k_1^2}W(k_1) \no \\
&&\times
\left[2\frac{k_z^2}{k^2}W(k)+2\frac{k_zk_{2z}}{k_2^2}W(k_2)-1\right]\ . 
\ea
Here, ${\bf k}_2=-{\bf k}_1-{\bf k}$. $B_3({\bf k}_1,{\bf
  k}_2,{\bf k}_3)$ is the real space matter bispectrum. 
We notice that the ensemble average of $B_3({\bf k}_1,{\bf
  k}_2,{\bf k}_3)$ is directly available from the same redshift surveys used
for RSD measurement. Since $\sum_i
{\bf k}_i=0$, ${\bf k}_1$, ${\bf k}_2$ and ${\bf k}_3$ lie in 
the same plane. Due to the isotropy of the universe, $B_3({\bf k}_1,{\bf
  k}_2,{\bf k}_3)$ does not depend on the inclination of the plane. So its
value is equal to the case where all ${\bf k}$ lie in the plane perpendicular to
the line of sight (namely the x-y plane). Redshift distortion does not
alter the $k_z=0$ Fourier mode.  Namely, $\delta^s({\bf k}_{\perp},
k_z=0)=\delta({\bf k}_{\perp},k_z=0)$.  This means that $B_3$ can be directly
measured from the observed Fourier mode with $k_{i,z}=0$.  Hence including
$C_{NG,3}$ in the 
calculation does not introduce extra fitting parameters and hence does not
weaken the cosmological constraints. 

The same trick does not apply to $C_{NG,j\geq 4}$. They  involve 4-th or
higher order correlation $\langle \delta({\bf 
  k}_1),\cdots, \delta({\bf k}_j)\rangle$, with ${\bf k}_1+\cdots+{\bf k}_j=0$.  In general,
${\bf k}_i$($i=1,2,3,4\cdots,j$) do not lie in the same plane. We are no
longer able to infer their values from the $k_z=0$ modes.  However, due to
extra suppression caused by $\tilde{W}<1$, we do not expect these terms to be
important. Nevertheless, the accuracy of neglecting these higher order terms
must be quantified through N-body simulations.

\subsection{A new formula on the redshift space power spectrum}
\label{subsec:RD}
We then propose the following formula for the redshift space power spectrum, 
\ba
\label{eqn:RD4}
P^s(k,u)&\simeq&
  \left\{P_{\delta\delta}(k)\left(1+f\tilde{W}(k)u^2\right)^2 \right.\\
&&\left. +u^4P_{\theta_S\theta_S}(k)+C_G(k,u)+C_{NG,3}(k,u)\right\}\no \\
&\times&
   D^{\rm FOG}_\delta(ku)D^{\rm FOG}_S(ku)\exp\left[-\frac{k^2u^2\sigma^2_{v_B}}{H^2}\right]\ . \no
\ea
In Eq. \ref{eqn:RD4}, $W(k)\equiv
f\tilde{W}(k)$ and $P_{\theta_S\theta_S}(k)$ are the cosmological information
we seek for.  We address one more
time that, {\it $C_G$ and $C_{NG,3}$  are not free
functions. They are uniquely determined by $W$.} This valuable
property is achieved by the proposed velocity decomposition and, in
particular, by the deterministic relation $\theta_\delta=\delta W$.   

There are  only very limited degrees of freedom in the FOG terms. Through
Eq. \ref{eqn:cumulantexpansion}, $D_\delta^{\rm 
  FOG}(ku)$ and $D_S^{\rm  FOG}(ku)$ may be well described by
$\sigma_{v_\delta}$, $\sigma_{v_S}$ and/or $K_{4,v_\delta}$, 
$K_{4,v_S}$, $K_{6,v_\delta}$,  $K_{6,v_S}$. Among them, $\sigma_{v_\delta}$ is determined by 
$W$ (Eq. \ref{eqn:sigmav}) and $\sigma_{v_S}$ is determined by
$P_{\theta_S\theta_S}$. So in principle neither
$\sigma_{v_\delta}$ nor $\sigma_{v_S}$ are free parameters \footnote{In
  reality, some $k$ modes of $W(k)$ and $P_{\theta_S\theta_S}(k)$  may not be
  well constrained to provide sufficently accurate prediction on
  $\sigma_{v_\delta}$ or $\sigma_{v_S}$. In this case, we may need to treat
  $\sigma_{v_\delta}$ and $\sigma_{v_S}$ as free parameters. }. 
Furthermore, since $\sigma_{v_B}^2\ll \sigma^2_{v_\delta}$
\cite{Pueblas09, Zheng12},  we may be able to set $\sigma_{v_B}=0$, depending on the
desired accuracy.

\subsubsection{Comparing to existing models}
Here we compare the proposed RSD formula (Eq. \ref{eqn:RD4}) with a few
existing RSD formulae. We choose these formulae because they can be compared
with ours relatively straightforwardly. So this comparison is by no means
complete. For brevity, we 
focus on the matter power spectrum for 
which we do not need to worry about the galaxy bias, especially the
nonlinear/non-deterministic 
bias. (1) The Kaiser formula plus the FOG effect (hereafter KF), 
\be
\label{eqn:KF}
P^s_{\delta\delta}(k,u)\simeq
P_{\delta\delta}(k)(1+fu^2)^2D^{\rm FOG}(ku)\ .
\ee
This is perhaps the most commonly used RSD formula. 
(2) The Desjacques \& Sheth 2010 formula (hereafter DS10, \cite{Desjacques10}),
\ba
\label{eqn:DS10}
P^s_{\delta\delta}(k,u)&=&P_{\delta\delta}(k)(1+fu^2)^2\no \\
&&\times \exp(-k_z^2\sigma_v^2/H^2)V_{\rm vir}(k_z) \ .
\ea
Here, $V_{\rm vir}$ is the damping caused by random motions of virialized
particles in halos.  $\sigma_v$ is the velocity dispersion of bulk motion. It
improves over the KF formula by correctly recognizing the two velocity
components (bulk motion and random motion). 
(3) The Scoccimarro 2004 formula (hereafter S04, \cite{Scoccimarro04}),
\ba
\label{eqn:S04}
P^s_{\delta\delta}(k,u)&=&\left(P_{\delta\delta}(k)+2u^2P_{\delta\theta}(k)+u^4P_{\theta\theta}(k)\right)\no \\
&&\times \exp(-k_z^2\sigma_v^2/H^2)\ . 
\ea
Here, $\sigma_v$ can be treated as the one calculated from the perturbation
theory or as
a free parameter. It improves over KF by taking into account differences in
the density and velocity nonlinear evolution. As a consequence, it does not
assume a deterministic relation between density and velocity. 
(4) The Taruya et al. 2010 formula (hereafter T10, \cite{Taruya10}),
\ba
\label{eqn:Taruya10}
P^s_{\delta\delta}(k,u)&=&\left(P_{\delta\delta}(k)+2u^2P_{\delta\theta}(k)+u^4P_{\theta\theta}(k)\right. \\
&+& \left. A(k,u)+B(k,u)\right)\exp(-k^2u^2\sigma_v^2/H^2)\ .  \no
\ea
The extra term $A$ involves integral over the bispectrum. The term B
involves convolution of two real space power spectra. 

Comparing to Eq. \ref{eqn:RD4}, we recognize a
number of corrections to existing formulae. Here we
briefly summarize the most 
significant ones. (1) Both KF and DS10 fail to capture the $\tilde{W}$ correction on $f$. So
  $f$ based on KF and DS10 is underestimated by a factor $\tilde{W}$. (2) S04
and T10 improve over KF and DS10 by correctly capturing the $\tilde{W}$
correction, although S04 and T10 do not make this correction explicitly. 
T10 further improves over S04 by including next order corrections ($A$ and
$B$).   In the appendix
\ref{sec:appendixB}, we prove that $A=C_{NG,3}$ in the limit ${\bf
  v}_S\rightarrow 0$. However, T10 adopted the approximation
$\epsilon_\delta=0$ in some intermediate steps.  so it  fails to capture some terms of the same
order as $A$ and $B$, as we do (Eq. \ref{eqn:CGreal4}).  More importantly,
calculating $A$ and $B$ requires heavy modeling or simulation
calibration. Our Eq. \ref{eqn:RD4} avoids these uncertainties. 
(3) The FOG effect is caused by three distinctly different velocity
components, so it may not
  follow a simple function form.  For this point, the closest match to our
  formula is DS10.  These complexities can bias the
  interpretation of the inferred $\sigma_v$ from RSD.

\subsection{Implications on the RSD cosmology}
\label{subsec:cosmology}
Our analysis above shows that what one can infer from RSD is the combination
$f\tilde{W}(k)$ instead of $f$.  Since $\tilde{W}(k)\sim
0.9$ at $k=0.1h/$Mpc and $z=0$, this can cause  $10\%$ underestimation in
$f$. We expect that it is at least a significant source causing the recently found
underestimation in $f$
\cite{Okumura11a,Jennings11a,Jennings11b,Kwan12,Bianchi12,delaTorre12}.
This systematical error overwhelms the $1\%$ level statistical accuracy in
$f$ by stage IV dark energy projects. With robust modeling of
$\tilde{W}$, it is promising to eliminate this systematical
error in the RSD cosmology. 

The  $\tilde{W}$ correction affects many applications of redshift
distortion. For example, \cite{Zhang07c} proposed an $E_G$ estimator combining
weak lensing and redshift distortion to test general relativity at
cosmological scale. This estimator is insensitive to galaxy bias and is also
less affected by initial fluctuations. \cite{Reyes10} made the first $E_G$
measurement and found $E_G=0.39\pm 0.06$ at an effective redshift $0.32$ and
$O(10) $Mpc$/h$ scale. This measurement confirms general relativity within
$\sim 20\%$  observational uncertainties.  This result has been used to
perform consistency tests of the $\Lambda$CDM cosmology and confirmed its
validity \cite{Lombriser11}. 
Given the existence of $\tilde{W}\neq 1$, the
expectation value of $E_G$  
should be corrected by a factor  $1/\tilde{W}$,  
\be
\label{eqn:EG}
E_G=\left[\frac{G_{\rm
      eff}}{G_{N}}\frac{\Omega_0}{f}\right]\frac{1}{\tilde{W}(k)}\ .
\ee
Future experiments such as BigBOSS+LSST, BigBOSS+Planck CMB lensing,
Euclid, SKA or
other combinations of spectroscopic surveys and imaging surveys are capable of
measuring $E_G$ to $1\%$ level statistical accuracy \cite{Zhang07c}. For these measurements, 
The $1/\tilde{W}$ correction is needed in order to correctly interpret the measured
$E_G$. 

\subsection{To do list}
Most statistics discussed in this section are too complicated to evaluate
analytically. In companion papers \cite{Zheng12}, we will use N-body simulations to
numerically evaluate  these statistics. An incomplete investigation list includes
\bi
\item The damping function $D^{\rm FOG}_\delta(ku)$,
  $D^{\rm FOG}_S(ku)$ and  $D^{\rm FOG}_B(ku)$. In particular, we will
  investigate the usage of the cumulant expansion theorem (Eq. \ref{eqn:cumulantexpansion}).  We will
  check if including $K_4$ and/or $K_6$ describes $D^{\rm FOG}_\delta(ku)$ and
  $D^{\rm FOG}_S(ku)$ accurately at scales of interest.  We will also check
  the accuracy of Eq. \ref{eqn:FOGB}. 
\item $\epsilon_\delta$, $\epsilon_S$ and $\epsilon_B$. In particular, we will
  quantify the accuracy of  Eq. \ref{eqn:DB} \& \ref{eqn:DR}.  
\item The accuracy of the proposed RSD formula (Eq. \ref{eqn:RD4}).  We will
  measure $C_{G,j}$ and $C_{NG,j}$ from simulations. We will also quantify the
  relative contribution of the leading terms, namely $C_{G,4}$ to $C_G$, 
  $C_{NG,3}$ to $C_{NG}$ and $P_{\theta_S\theta_S}$ with respect to $C_S$. We
  will seek the possibility to further improve the RSD modeling, if needed. 
\ei

\section{Reconstruction of the 3D velocity field in redshift surveys}
\label{sec:3D}
So far we focus on  inferring the statistical average of the velocity field
including $W\propto P_{\delta\theta}$, $P_{v_Sv_S}$, $\sigma_{v_\delta}$,
$\sigma_{v_S}$ and $\sigma_{v_B}$.  
Can we go a step further to construct the full 3D velocity field? The gain would be
huge since the 3D velocity field contains much more information and
has much more applications. One of such applications is the kinetic Sunyaev
Zel'dovich (kSZ) tomography \cite{Ho09,Shao11b}.

The diffuse kinetic Sunyaev Zel'dovich (kSZ) effect \cite{SZ80} caused by
the intergalactic  medium is a potentially powerful probe for missing
baryons. Unfortunately, measuring kSZ is very difficult, due to the weakness
of the kSZ signal, the lack of spectral feature and various overwhelming
contaminations such as primary CMB, the thermal SZ effect and cosmic infrared
background. The state of art experiments such as ACT  \cite{Dunkley11,Hand12}
and SPT \cite{Shirokoff11,Reichardt12} has reported the
first detection of the kSZ effect of galaxy clusters
\cite{Hand12}. Nevertheless,  the diffuse kSZ  is still elusive.

The measurement can  be 
revolutionized by the kinetic SZ tomography
\cite{Ho09,Shao11b}. The key is to reconstruct the galaxy velocity field and 
then construct the galaxy momentum field. We then correlate the projected
momentum field  with CMB to measure the kSZ effect. This kSZ tomography automatically
eliminating all contaminations of scalar type. A   combination of BigBOSS and Planck  
is promising to measure kSZ to  better than $10\sigma$ at a number of 
redshift bins \cite{Shao11b}. 

\subsection{Proposals on the velocity reconstruction} 
How to reconstruct the 3D velocity field? If we can, which velocity component
can be reconstructed? There are numerous works over a long history. We are not
at a stage to overview these works. Rather, we outline our approach, based
upon the proposed velocity decomposition. 

Since observationally we only have the density field,
which is a scalar field, the information budget does not allow us to 
reconstruct all the three ${\bf v}_\delta$,
${\bf v}_S$ and ${\bf v}_B$ vector fields. However, since ${\bf v}_\delta$ is completely
correlated with the density field, it is promising to reconstruct ${\bf v}_\delta$
from the observed density field.  Eq. \ref{eqn:thetam}
guides us to propose a linear estimator for the 3D peculiar velocity reconstruction
through the 3D density field.  Since we observe the redshift space
overdensity $\delta^s$ instead, this linear estimator should operate on
$\delta^s({\bf k})$,
\be
\label{eqn:estimator}
\hat{\theta}_\delta({\bf k})=\delta^s({\bf k})\hat{W}^s({\bf k})\ .
\ee
We use the superscript ``$s$'' to denote  properties in redshift
space. $\tilde{W}^s$ corresponds to an anisotropic window function operating
on the anistropic density field in redshift space. Now the reconstruction of
the stochastic 3D velocity filed is 
simplified into the reconstruction of a deterministic function $W^s$.  Due to
the axial symmetry along the light of sight, $W^s({\bf k})=W^s(k,u)$. So it
only has 2D degrees of 
freedom. These degrees of freedom are further limited by the asymptotic
behaviors of $W^s(k,u)$ at $ku\rightarrow 0$ and at $k\rightarrow \infty$. 

We want the peculiar velocity estimator (Eq. \ref{eqn:estimator}) to have no
multiplicative error. This requires 
\be
\label{eqn:Ws}
\hat{W}^s({\bf k})=\frac{W(k)}{C_v({\bf k})}=\frac{W(k)}{r_{\delta\delta^s}({\bf
  k})}\sqrt{\frac{P_{\delta\delta}(k)}{P^s_{\delta\delta}({\bf k})}}\ .
\ee
$W^s$ differs from the real space one by a
direction dependent  factor $1/C_v({\bf k})$. This factors arises because the
redshift space density 
$\delta^s$ is not completely correlated with the real space density
$\delta$. To better understand this point and to derive Eq. \ref{eqn:Ws}, we carry out a
decomposition of  $\delta^s$  into two parts, 
\be
\label{eqn:deltasdecomposition}
\delta^s=\delta^s_v+\delta^s_S\ .
\ee
 We
require $\delta^s_v$ to be completely correlated with the underlying
overdensity $\delta$ and hence to $\theta_\delta$. This is the part that we can
use to recover the peculiar velocity. For this reason, we label this part with a subscript
``$v$''. $\delta^s_S$ is uncorrelated with $\delta$ and $\theta_\delta$. It
causes the stochasticity in $\delta$-$\delta^s$ relation and 
contaminates the velocity reconstruction.  So it is labelled with a subscript
``S''. Since $\delta_v$ is completely
correlated with $\delta$, 
\be
\delta^s_v({\bf k})=\delta({\bf k}) C_v({\bf k})\ ,
\ee
with the deterministic function $C_v$ to be determined. Through the relation
$\langle \delta({\bf k})\delta^s({\bf k}^{'})\rangle=\langle
\delta({\bf k})\delta^s_v({\bf k}^{'})\rangle =\langle
\delta({\bf k})\delta({\bf k}^{'})\rangle C_v({\bf k})$,
we obtain 
\be
C_v({\bf k})=\frac{P_{\delta\delta^s}({\bf k})}{P_{\delta\delta}(k)}=r_{\delta\delta^s}({\bf
  k})\sqrt{\frac{P^s_{\delta\delta}(k,u)}{P_{\delta\delta}(k)}}\ .
\ee
Here, $P_{\delta\delta^s}$  is the cross power spectrum between $\delta$ and  $\delta^s$.
$r_{\delta\delta^s}$ is the corresponding cross correlation coefficient. It has the
asymptotic behavior $r_{\delta\delta^s}\rightarrow 1$ when $ku\rightarrow 0$ and
$r_{\delta\delta^s}\rightarrow 0$ when $ku\rightarrow \infty$. Notice that $r_{\delta\delta^s}$ depends
on both $k$ and $u$. So do $C_v$ and $W^s$. Hereafter we will write them as
$r_{\delta\delta^s}(k,u)$, $W^s(k,u)$ and $C_v(k,u)$ to highlight these dependences.

Both $P_{\delta\delta}(k)$ and $P^s_{\delta\delta}(k,u)$ are observables. To
evaluate  $\hat{W}^s$ requires just one extra input, $r_{\delta\delta^s}(k,u)$. $r_{\delta\delta^s}$ has a well defined asymptotic behavior $r_{\delta\delta^s}\rightarrow
1$ when $ku\rightarrow 0$ and $r_{\delta\delta^s}\rightarrow 0$ when $|ku|\rightarrow
\infty$. In the appendix \ref{sec:appendixC}, we show that $r$ is
uniquely fixed by $W$ and
other quantities which can be inferred from $P_{\delta\delta}^s(k,u)$.  Hence
the observed RSD allows us to  figure out $W^s$ and then carry out the velocity
reconstruction.   In this sense,  {\it RSD is a source of  information
crucial for the velocity reconstruction,
instead of source of noise as in many other apprroaches of velocity
reconstruction}. 

In the appendix \ref{sec:appendixD}, we propose another approach to
reconstruct the 3D velocity field. It shows more clearly the crucial role of RSD in the velocity
reconstruction. 

\subsection{Reconstruction errors and remodies}
A correct $W^s(k,u)$ allows to reconstruct ${\bf v}_\delta$ free of multiplicative
error. Unfortunately, additive errors persist, due to the stochastic 
component $\delta^s_S$.  The reconstructed velocity divergence is
\ba
\hat{\theta}_\delta({\bf k})&=&\theta_\delta({\bf k})+\theta^s_S({\bf k}) \ , \\
\theta^s_S({\bf k})&\equiv& \delta^s_S({\bf k})\hat{W}^s(k,u) \ . \no 
\ea
The additive error  in the reconstructed velocity is then 
\be
\label{eqn:verror}
v^s_S({\bf k})=\frac{iH\theta^s_S({\bf k}) {\bf k}}{k^2}\ .
\ee
If we measure the velocity power spectrum directly through the reconstructed
velocity field, it will suffer from an additive error.  So 
cosmological applications of the directly measured auto power spectrum can be very
limited, if any. Fortunately, one can circumvent this problem
straightforwardly.

We start with a valuable property, that the cross power spectrum between the
reconstructed velocity and  the density
distribution is  unbiased,
\be
\hat{P}_{\delta \theta}=P_{\delta \theta}=P_{\delta\theta_\delta}\ .
\ee
For the same reason, the additive error does not bias the kSZ tomography
\cite{Ho09,Shao11b} and
hence does not bias the effort to search for missing baryons. 

Since $\theta_\delta$ is completely correlated with the matter density, we can
measure the auto power spectrum, through the relation
\be
P_{\theta_\delta\theta_\delta}=\frac{P_{\delta\theta_\delta}^2}{P_{\delta\delta}}\ .
\ee
The auto power spectrum measured in this way is free of additive error
discussed above.

\subsection{To do list}
Through N-body simulations, a number of key issues will be investigated in
future works,
\bi
\item $r_{\delta\delta^s}(k,u)$, $W^s(k,u)$ and $C_v(k,u)$. We are then able to quantify  the
 additive error in the reconstructed velocity.  Furthermore, we want to understand the origin of
  $\delta^s_S$, which causes the stochasticity in the $\delta$-$\delta^s$
  relation and degrades the reconstruction performance. 
\item The accuracy of Eq. \ref{eqn:rrs} to model $r_{\delta\delta^s}$. 
\item The accuracy of the velocity reconstruction. It is determined by the
  accuracy of the inferred $W$ from the observed RSD and the accuracy of the
  modeled $r_{\delta\delta^s}$. Measurement errors  in the galaxy distribution 
  such as shot noise further complicates the reconstruction.  We will take
  them into account for more realistic quantification of the reconstruction
  performance. 
\item Further investigation of the velocity reconstruction approach proposed
  in the appendix \ref{sec:appendixD}. 
\ei 

\section{From matter distribution to galaxy distribution}
\label{sec:galaxy}
In reality, we have 3D galaxy distribution instead of 3D matter
distribution. The matter-galaxy relation is complicated. Nevertheless, the
above methods to model the matter redshift space 
distortion and to carry out velocity reconstruction can be extended to the 3D
galaxy distribution straightforwardly and robustly. The procedure is as
follows. 

\subsection{The redshift space galaxy power spectrum}
First,  we decompose the galaxy velocity ${\bf v}_g$ into a irrotational part
${\bf v}_{E,g}$ and a rotational part ${\bf v}_{B,g}$. For clarity, we use
the subscript ``g'' to denote galaxy properties.  This step is essentially the
same as the case of the matter field. 

But the next step is different.  Now we need to  decompose the irrotional galaxy 
velocity ${\bf v}_{E,g}$ into two eigen-modes such that one part (${\bf
  v}_{\delta_g}$) is completely 
correlated with the galaxy overdensity $\delta_g$ and the other part (${\bf
  v}_{S,g}$) is completely uncorrelated with $\delta_g$.  Following \S
\ref{sec:decomposition},  we obtain the $\theta_{\delta_g}$-$\delta_g$ relation,
\be
\label{eqn:thetag}
\theta_{\delta_g}({\bf k})=W_g(k) \delta_g({\bf k})\ .
\ee
Here $\theta_{\delta_g}$ is the divergence of ${\bf v}_{\delta_g}$. The window
function $W_g$ is given by 
\be
\label{eqn:Wg}
W_g(k)\equiv \frac{P_{\delta_g\theta_g}(k)}{P_{\delta_g\delta_g}(k)}\  .
\ee
We recognize that $W_g(k\rightarrow
0)=f/b_g(k\rightarrow 0)=\beta$. Here, $b_g(k\rightarrow 0)$
is the linear galaxy bias. We have assumed no galaxy 
velocity bias at sufficiently large scale.
We also define 
\be
\label{eqn:tildeWg}
\tilde{W}_g(k)\equiv \frac{W_g(k)}{W_g(k\rightarrow
  0)}=\frac{W_g(k)}{\beta}=\frac{1}{\beta}\frac{P_{\delta_g\theta_g}}{P_{\delta_g\delta_g}}\ .
\ee
One can compare
Eq. \ref{eqn:thetag}, \ref{eqn:Wg}, \ref{eqn:tildeWg} with 
Eq. \ref{eqn:thetam}, \ref{eqn:W}, \ref{eqn:tildeW}, respectively. There are
important differences. For example,  $W$ and $W_g$ differ by a factor $b_g$ at
large scale. Consequently, the relation $W=f\tilde{W}$ is replaced by
$W_g=\beta\tilde{W}_g$. 

The final step is to replace quantities of the matter field in
previous sections by corresponding galaxy quantities to obtain results
applicable to the galaxy field.   The transformation is,
\ba
\label{eqn:transformation}
\delta\rightarrow \delta_g\ ,\  \theta\rightarrow \theta_g\ ,\ W\rightarrow
W_g\ ,\ \tilde{W}\rightarrow \tilde{W}_g\ ,\ f\rightarrow \beta\ ,\\ 
{\bf v}\rightarrow {\bf v}_g\ ,\ {\bf v}_\delta \rightarrow {\bf
  v}_{\delta_g}\ ,\ {\bf v}_S \rightarrow {\bf v}_{S,g} \ ,\  {\bf v}_B \rightarrow {\bf
  v}_{B,g}\ ,\no \\
D_{\delta}\rightarrow D_{\delta_g}\ ,\ D_S\rightarrow
D_{S,g}\ ,\ D_B\rightarrow D_{B,g}\ ,\ \cdots \no
\ea

We explicitly show one example.  In \S \ref{subsec:RD}, we suggest to use 
Eq. \ref{eqn:RD4} to model the matter power spectrum in redshift space. 
Applying the above transformation (Eq. \ref{eqn:transformation}) to
Eq. \ref{eqn:RD4}, we obtain the formula for the redshift space galaxy power
spectrum, 
\ba
\label{eqn:RD4g}
P^s_{\delta_g\delta_g}(k,u)&\simeq &\left\{P_{\delta_g\delta_g}(k)(1+\beta
\tilde{W}_g(k)u^2)^2 \right.\\
&+&\left. u^4P_{\theta_{S,g}\theta_{S,g}}(k)+C_{G,g}(k,u)+C_{NG,3g}(k,u)\right\}\no \\
&\times & D^{\rm FOG}_{\delta_g}(ku)D^{\rm
  FOG}_{S,g}(ku)\exp\left[-\frac{k^2u^2\sigma^2_{v_{B,g}^2}}{H^2}\right] \ . \no
\ea
Here, $C_{NG,3g}$ and  $C_{G,g}$ are defined by applying the transformation in
Eq. \ref{eqn:transformation} to Eq. \ref{eqn:CNGreal} and \ref{eqn:CGreal}.
In the limit $k\rightarrow 0$, we recover the correct behavior 
$P_{\delta_g\delta_g}^s\rightarrow P_{\delta_g\delta_g}(1+\beta u^2)^2$.

\subsection{On the interpretation of $E_G$}
The above transformation is in general straightforward to apply. Nevertheless,
extra care should be given for some special cases. For example, it turns out that the
$\tilde{W}$ in the $E_G$  
estimator (Eq. \ref{eqn:EG}) may still be interpreted as that of the
matter field instead of the galaxy field, for the complexity that  it
involves not only the galaxy field (redshift distortion), but also the
matter field (weak lensing) and the galaxy field.

By construction, $E_G$ is proportional to the ratio of galaxy-lensing cross correlation and
galaxy-velocity cross correlation \cite{Zhang07c}. Namely $E_G\propto
P_{\delta_g\delta}/P_{\delta_g\theta_g}$.  Notice that $\delta$ is the matter
overdensity. In the large scale limit ${\bf v}_{\delta_g}\rightarrow {\bf
  v}_\delta$ ($\theta_{\delta_g}\rightarrow \theta_\delta=\delta W$),
we have $E_G\propto 1/W=1/(f\tilde{W})$, even for the galaxy field.  This property is desirable since the correction term
$\tilde{W}$  can be robustly quantified through N-body simulations, free of
uncertainties in modeling galaxy formation. Nevertheless, we shall use
simulations to quantify the accuracy of the key approximation $\theta_{\delta_g}\simeq \theta_\delta$
and the scale where this approximation breaks.

%Assuming no velocity bias in the velocity component ${\bf
%  v}_\delta$ (namely ${\bf v}_{\delta_g}={\bf
%  v}_\delta$)\footnote{This is a key issue to investigate in order to
%  robustly quantify $E_G$.}, we have
%$\theta_{\delta_g}=\theta_\delta=\delta W$.
%To eliminate $\delta$ in the numerator,  we should decompose the
%galaxy velocity with respect to the matter 
%field, instead of the galaxy number density field. We then have
%$\theta_g=\delta W+\theta_{S,g}$. Here we have assumed no velocity bias in the
%velocity component ${\bf v}_\delta$. 
%Under reasonable assumptions,  $P_{\delta_g\theta_{S,g}}=0$ \footnote{Strictly
%  speaking, the decomposition only guarantees $\langle
%  \delta\theta_{S,g}\rangle=0$, but not $\langle \delta_g\theta_{S,g}=0$. To reach $P_{\delta_g\th%eta_{S,g}}=0$ requires  that a velocity
%  component uncorrelated with the matter density field should be uncorrelated
%  with the galaxy distribution.  This is an issue to be investigated in future
%works. }. 
%We then have 
%$P_{\delta_g\delta}/P_{\delta_g\theta_g}=P_{\delta_g\delta}/(WP_{\delta_g\delta})\propto
%1/(f\tilde{W})$. 

\subsection{Velocity reconstruction through 3D galaxy distribution}
Following the case of velocity reconstruction through 3D redshift space matter distribution,
we derive the linear estimator to reconstruct ${\bf v}_{\delta_g}$ from 3D redshift
space galaxy overdensity $\delta_g^s$, 
\be
\label{eqn:estimatorg}
\hat{\theta}_{\delta_g}({\bf k})=\delta_g^s({\bf k})\hat{W}_g^s({\bf k})\ .
\ee
The anisotropic window function $\hat{W}_g^s({\bf k})$ is derived to be 
\be
\label{eqn:Wsg}
\hat{W}_g^s({\bf k})=\frac{W_g(k)}{C_{v,g}({\bf k})}=\frac{W_g(k)}{r_{\delta_g\delta^s_g}({\bf
  k})}\sqrt{\frac{P_{\delta_g\delta_g}(k)}{P^s_{\delta_g\delta_g}({\bf k})}}\ .
\ee
Here, $r_{\delta_g\delta^s_g}$ is the cross correlation coefficient between
the real space $\delta_g$ and the redshift space $\delta_g^s$.  It can be
computed by applying the transformation Eq. \ref{eqn:transformation} to
Eq. \ref{eqn:rrs}.  One can
compare between Eq. \ref{eqn:estimator} \& \ref{eqn:estimatorg} and between
Eq. \ref{eqn:Ws} \& \ref{eqn:Wsg} for similiarities and differences.

Our velocity reconstruction method
significantly alleviates the problem of galaxy bias in some existing
velocity reconstruction methods. Even better, in the limit of a deterministic bias
($\delta_g=b_g\delta$), it completely overcomes it. 
Since $W_g\propto 1/b_g$, the reconstructed 3D velocity is $\propto
W_g\delta_g\propto b_g^0$, independent of $b_g$. This is also the case for the
inferred velocity power  spectrum, 
\be
P_{\theta_{\delta_g}\theta_{\delta_g}}(k)=\frac{P_{\delta_g\theta_{\delta_g}}^2(k)}{P_{\delta_g\delta_g}(k)}\propto
b_g^0\ .
\ee

In future works we will redo the numerical analysis of the matter density field for the
halo number density field, through  N-body simulations of $O(10) $ Gpc$^3$
volume in total. We will then proceed to mock catalogs of
galaxies. Eventually we plan to develop efficient and sophisticated codes applicable to real
data of spectroscopic redshift surveys.

\section{Discussions and summary}
We have laid out the methodology to carry out 3D velocity reconstruction from
3D matter and galaxy distribution. The method is based upon a velocity
decomposition into three eigen-modes with physical motivation and of mathematical
uniqueness. The same decomposition also helps us to derive a potentially more
robust RSD  formula. Through it we find that the 
inferred structure growth rate based upon some simplified versions of RSD
modeling can be severely underestimated. In a series of companion papers
\cite{Zheng12} we will analyze N-body simulations to measure statistics of the
three velocity eigen-modes, to test the accuracy of the proposed RSD formula
and to quantify the performance of the proposed velocity reconstruction. 

\section*{Acknowledgments}
This work was supported by the national science foundation of China
(grant No. 11025316 \& 11121062 and 10873035 \& 11133003), National Basic
Research Program of China (973 Program) under grant No.2009CB24901 and  the
CAS/SAFEA 
International Partnership Program for  Creative Research Teams (KJCX2-YW-T23).

\appendix
\section{Calculating the ${\bf v}_\delta$ induced RSD}
\label{sec:appendixA}
Throughout this section we only deal with the velocity component ${\bf v}_\delta$. Hence
for brevity we neglect the subscript ``$\delta$'' and denote $v_{1z,\delta}=v_1$,
$v_{2z,\delta}=v_2$, $\lambda_{1,2}\equiv ik_zv_{1,2}/H$ and
$\lambda_{\delta}\equiv \lambda_1-\lambda_2$.  We will adopt two tricks in
\cite{Scoccimarro04,Taruya10} to faciliate the derivation. The first is the relation 
$\langle \delta_1\exp\lambda_\delta\rangle=\frac{\partial}{\partial
  a_1}\langle \exp(\lambda_\delta+a_1\delta_1)\rangle_{a_1=0}$. In combination
with the cumulant
expansion theorem $\langle \exp X\rangle=\exp(\sum \langle
  X^n\rangle_c/n!)$,  we have
\ba
\langle \delta_1\exp\lambda_\delta\rangle&=&\langle \exp\lambda_\delta\rangle
\sum_{n\geq 1} \frac{n\langle \lambda_\delta^{n-1}\delta_1\rangle_c}{n!}\no \\
&=&\langle \exp\lambda_\delta\rangle
\sum_{n\geq 0} \frac{\langle
  \lambda_\delta^{n}\delta_1\rangle_c}{n!}\no \\
&=&\langle \exp\lambda_\delta\rangle
\sum_{n\geq 0} \frac{\langle
  \lambda_\delta^{n}\delta_1\rangle_c}{n!}\ .
\ea
The second relation is $\langle
\delta_1\delta_2\exp\lambda_\delta\rangle=\frac{\partial^2}{\partial 
  a_1\partial a_2}\langle
\exp(\lambda_\delta+a_1\delta_1+a_2\delta_2)\rangle_{a_1=0,a_2=0}$. Again with
the cumulant expansion theorem, we have
\ba
\langle \delta_1\delta_2\exp\lambda_\delta\rangle=\langle
\exp\lambda_\delta\rangle\times \\
\left[\sum_{j\geq 1} \frac{\langle \lambda_\delta^j\delta_1\rangle_c}{j!}\sum_{n\geq 1} \frac{\langle
    \lambda^n\delta_2\rangle_c}{n!}+\sum_{n\geq 0} \frac{\langle
    \delta_1\delta_2\lambda_\delta^n\rangle_c}{n!}  \right]\ .\no 
\ea
Putting all pieces together, we obtain 
\ba
\label{eqn:A3}
\langle (1+\delta_1)(1+\delta_2)\exp\lambda_\delta\rangle=D^{\rm
  FOG}_\delta(k_z)(1+\epsilon_\delta({\bf r},k_z)) \no \\ \left(1+\sum_{n\geq 1} \frac{\langle
  \lambda_\delta^{n}(\delta_1+\delta_2)\rangle_c}{n!} +\sum_{n\geq 0}
\frac{\langle \delta_1\delta_2\lambda_\delta^n\rangle_c}{n!} \right.\\
+\left. \sum_{j\geq 1} \frac{\langle \lambda_\delta^j\delta_1\rangle_c}{j!}\sum_{n\geq 1} \frac{\langle
    \lambda_\delta^n\delta_2\rangle_c}{n!}\right)\ . \no
\ea
 Through the cumulant expansion
theorem, we  have  
\ba
\label{eqn:epsilondelta}
1+\epsilon_\delta({\bf r},k_z)&=&\exp\left[\sum_{\alpha+\beta\geq 2}
  \frac{\langle \lambda_\delta^{\alpha+\beta}\rangle_c}{(\alpha+\beta)!}-\frac{\langle\lambda_1^\alpha\rangle_c}{\alpha!}
\frac{\langle-\lambda_2^\beta\rangle_c}{\beta!}\right]\no \\
&=&\exp\left[\frac{k_z^2\langle
    v_1v_2\rangle}{H^2}+O(v^4)\right] \ .
\ea 
Here, terms odd in the power of $v$ vanish. For example, due
to symmetry of $v\leftrightarrow -v$, we have
$\langle (v_1-v_2)^3\rangle=0$, $\langle v_1^3\rangle=\langle
v_2^3\rangle$, and hence $\langle v_1v_2^2\rangle-\langle
v_1^2v_2\rangle=0$.For this reason, the next leading order terms in
Eq. \ref{eqn:epsilondelta} is of the order $v^4$, whose exact
expression can be obtained following Eq. \ref{eqn:DR}.

Collecting terms of the same orders together, we have 
\ba
\label{eqn:RSDm}
\langle
(1+\delta_1)(1+\delta_2)\exp\lambda_\delta\rangle=D^{\rm FOG}_\delta(k_z)
\no \\
\times  \left[1+\langle \delta_1\delta_2\rangle+i\frac{k_z}{H}\langle
   \delta_2v_1-\delta_1v_2\rangle+\frac{k_z^2}{H^2}\langle
   v_1v_2\rangle\right.\\\left.+C_{NG}({\bf r},k_z)+C_{G}({\bf r},k_z)
   \right] \ . \no
\ea
Two correction terms $C_G$ and $C_{NG}$ show up.  Both arise from the
nonlinear real space-redshift space mapping. But $C_G$ exhausts all  high order
corrections if the density and velocity fields are Gaussian. So we denote it with a
subscript ``G''.  It has an exact analytical expression, 
\ba
\label{eqn:CGreal}
C_G({\bf r},k_z)&=&\left[\exp\left(\frac{k_z^2\langle v_1v_2\rangle}{H^2}\right)-1\right] \no\\
&\times&\left(\langle
   \delta_1\delta_2\rangle+i\frac{k_z}{H}\langle
   \delta_2v_1-\delta_1v_2\rangle\right)\\
&+&\frac{k_z^2}{H^2}\langle
   \delta_1v_2\rangle\langle\delta_2v_1\rangle \exp\left(\frac{k_z^2\langle
     v_1v_2\rangle}{H^2}\right)\no \\ 
&+&\left[\exp\left(\frac{k_z^2\langle
       v_1v_2\rangle}{H^2}\right)-\frac{k_z^2}{H^2}\langle v_1v_2-1\rangle\right] \ . \no
\ea
We notice that this analytical result for Gaussian field has been derived by
\cite{Scoccimarro04} and shown as their Eq. 32. 

As shown in \S \ref{sec:RD}, $C_G$ can be robustly calculated combining
observations without knowing the underlying cosmology and without introducing
extra unknown parameters. The leading order term is 4-th power in the density
field, with 
\ba
\label{eqn:CGreal4}
C_{G,4}({\bf r},k_z) &=& \frac{k_z^2\langle v_1v_2\rangle}{H^2}\left(\langle
   \delta_1\delta_2\rangle+i\frac{k_z}{H}\langle
   (v_1-v_2)(\delta_1+\delta_2)\rangle\right)\no \\
&+&\frac{k_z^2}{H^2}\langle
   \delta_1v_2\rangle\langle\delta_2v_1\rangle+\frac{1}{2}\frac{k_z^4}{H^4}\langle
   v_1v_2\rangle^2 \ .
\ea

$C_{NG}$ exhausts all high order corrections arising from
non-Gaussianities in the density and veocity fields. For this
reason, we denote it with the  subscript ``NG''. It is the sum of an infinite
series of $j$-th order 
correlations with $j\geq 3$.  
\ba
\label{eqn:CNGreal}
C_{NG}&=&\sum_{j\geq 3} C_{NG,j}({\bf r},k_z) \\
&=&\left[i\frac{k_z\langle
   \delta_1\delta_2(v_1-v_2)\rangle_c}{H} -\frac{k_z^2\langle
     (v_1-v_2)^2(\delta_1+\delta_2)\rangle_c}{2H^2}\right]\no\\
&&+\cdots  \ . \no
\ea
In the above equation, we only show the explicit expression of
$C_{NG,j=3}$. 

So far the results are exact. However, to realistically evaluate
$C_{NG}$, we need to truncate somewhere in the $C_{NG,j}$ series. We
argue that, due to extra suppression 
$\tilde{W}\ll 1$ in the deeply nonlinear region, $j>3$ terms should be
smaller than $C_{NG,j=3}$. So it may be reasonably accurate to keep only
$C_{NG,j=3}$. Nevertheless, we will check this
approximation through N-body simulations and investigate if 
$C_{NG,j\geq 4}$ should be included in the calculation. 

In the appendix \ref{sec:appendixB}, we will prove that $C_{NG,j=3}$ is
equivalent to the term $A$ in \cite{Taruya10}. Strictly speaking, its Fourier
transform $C_{NG,j=3}(k,u)=A(k,u)$ in the limit of ${\bf
  v}_S\rightarrow 0$.

\section{The $C_{NG,3}$-$A$ relation}
\label{sec:appendixB}
The term $A$ is derived by \cite{Taruya10} as an additive correction to the Kaiser
formula. \cite{Taruya10} does not distinguish between ${\bf v}_\delta$ and ${\bf
  v}_S$, so the velocity showing up in the expression of $A$ is ${\bf
  v}_E={\bf v}_\delta+{\bf v}_S$. We consider the limit ${\bf
  v}_S=0$. Re-expressed in our notations, $A$ in 
\cite{Taruya10} is 
\ba
\label{eqn:A}
A(k,u)&=&i\frac{k_z}{H}\int  \langle
(v_1-v_2)(\delta_1-\frac{\nabla_1v_1}{H})(\delta_2-\frac{\nabla_2v_2}{H})\rangle\no
\\
&& \exp(-i{\bf
  k}\cdot {\bf r}) d^3{\bf r} \\
&=&ik_z\int  \langle (v_1-v_2)\delta_1\delta_2\rangle \exp(-i{\bf
  k}\cdot ({\bf r}) d^3{\bf r}\no
\\
&+& i\frac{k_z}{H^2}\int  \langle
(v_2\delta_2\nabla_1v_1-v_1\delta_1\nabla_2v_2+\delta_1v_2\nabla_2v_2\no \\
&&-\delta_2v_1\nabla_1v_1)\rangle \exp(i{\bf
  k}\cdot ({\bf x}_1-{\bf x}_2)) d^3{\bf x}_1d^3{\bf x}_2 \frac{1}{V} \no
\\
&+&i\frac{k_z}{H^3}\langle (v_1-v_2)\nabla_1v_1\nabla_2v_2\rangle\no \\
&& \exp(i{\bf
  k}\cdot ({\bf x}_1-{\bf x}_2)) d^3{\bf x}_1d^3{\bf x}_2 \frac{1}{V}\ . \no
\ea
Here $V$ is the total volume. 
Since  $v_2\delta_2\nabla_1v_1 \exp(i{\bf k}\cdot{\bf
  x}_1)=\nabla_1[v_2\delta_2v_1\exp(i{\bf k}\cdot{\bf
  x}_1)]-ik_zv_2\delta_2v_1 \exp(i{\bf k}\cdot{\bf
  x}_1)$ and since $\nabla_1 (\cdots)$ integrates to zero, the term
$v_2\delta_2\nabla_1v_1$ in the above equation can be replaced by the term
$-ik_zv_2\delta_2v_1$. In total we can do the following replacements in
Eq. \ref{eqn:A}, 
\ba
v_2\delta_2\nabla_1v_1 &\rightarrow& -ik_zv_2\delta_2v_1\ ,\no \\
v_1\delta_1\nabla_2v_2 &\rightarrow& ik_zv_1\delta_1v_2 \ ,\no \\
\delta_1v_2\nabla_2v_2 &\rightarrow& ik_z\frac{1}{2} \delta_1v_2^2\ , \no\\
\delta_2v_1\nabla_1v_1 &\rightarrow& -ik_z\frac{1}{2}\delta_2v_1^2\ ,\no \\
v_1\nabla_1v_1\nabla_2v_2 &\rightarrow &k_z^2\frac{1}{2}v_1^2v_2\ ,\no \\
v_2\nabla_1v_1\nabla_2v_2 &\rightarrow &k_z^2\frac{1}{2}v_2^2v_1 
\ea
Comparing to Eq. \ref{eqn:CGreal4}, we prove that, in the limit ${\bf
  v}_S\rightarrow 0$, $C_{NG,3}(k,u)= A(k,u)$. 

On the other hand, our $C_{G,4}$ is not equal to the $B$ term in \cite{Taruya10}, due to
differences in the methods and differences in approximations made. For
example,  \cite{Taruya10} sets $\epsilon_\delta=0$ in their Eq. 18. Inclusion
of $\epsilon_\delta\neq 0$ in our derivation 
brings up new terms such as  the term $\langle v_1v_2\rangle^2$ in
Eq. \ref{eqn:CNGreal}. In future works we will test against N-body simulations
to compare the two results. 

\section{Modeling $r_{\delta\delta^s}(k,u)$}
\label{sec:appendixC}
Following the derivation of $P^s_{\delta\delta}$, the real space
-redshift space density cross power spectrum $P_{\delta\delta^s}$ is given by 
\ba
P_{\delta\delta^s}(k,u)&=&\int \left\langle
(1+\delta_1)(1+\delta_2)\exp[ik_zv_{1z}/H]\right\rangle \\
&& \exp(-i{\bf k}\cdot{\bf
  r})d^3{\bf r} \no \\
&=&\int \left\langle
(1+\delta_1)(1+\delta_2)\exp\left(i\frac{k_zv_{1\delta,z}}{H}\right)\right\rangle
\no \\
&& \exp(-i{\bf k}\cdot{\bf
  r})d^3{\bf r} \sqrt{D^{\rm FOG}_S(k_z)D^{\rm FOG}_B(k_z)}\ . \no
\ea
Replacing $\lambda_\delta$ in Eq. \ref{eqn:A3} with
$\lambda_1$, we obtain
\ba
\langle (1+\delta_1)(1+\delta_2)\exp\lambda_1\rangle=\langle
\exp\lambda_1\rangle \no \\ \left(1+\sum_{n\geq 1} \frac{\langle
  \lambda_1^{n}(\delta_1+\delta_2)\rangle_c}{n!} +\sum_{n\geq 0}
\frac{\langle \delta_1\delta_2\lambda_1^n\rangle_c}{n!} \right.\\
+\left. \sum_{j\geq 1} \frac{\langle \lambda_1^j\delta_1\rangle_c}{j!}\sum_{n\geq 1} \frac{\langle
    \lambda_1^n\delta_2\rangle_c}{n!}\right)\ . \no
\ea
We then have
\ba
P_{\delta\delta^s}(k,u)&=&\left[P_{\delta\delta}(k)(1+f\tilde{W}(k)u^2)+C^{rs}_{NG}(k,u)\right] \\
&&\times  \sqrt{D^{\rm
    FOG}_\delta(k_z) D^{\rm
    FOG}_S(k_z)D^{\rm FOG}_B(ku)}\ . \no
\ea
The high order correction term 
\ba
C^{rs}_{NG}(k,u)&=&y\times P_{\delta\delta}(k)f\tilde{W}u^2+\int  \exp(-i{\bf k}\cdot{\bf
  r})d^3{\bf r}  \\
&&\times \left[\sum_{n\geq 1}
\frac{\langle \delta_1\delta_2\lambda_1^n\rangle_c}{n!}+\sum_{n\geq 2} \frac{\langle
    \lambda_1^n\delta_2\rangle_c}{n!}(1+y)\right]\ . \no
\ea
Here, $y\equiv \sum_{j\geq 1} \langle
\lambda_1^j\delta_1\rangle_c/j!=\sum_{j\geq 2} \langle
\lambda_1^j\delta_1\rangle_c/j!$. We can resum $C_{NG}^{rs}$  in order of the power of
$\delta$,
\ba
C^{rs}_{NG}(k,u)&=&\sum_{j\geq 3} C^{rs}_{NG,j}(k,u)\\
&=&\int  e^{-i{\bf k}\cdot{\bf
  r}}d^3{\bf r} \left[\langle \delta_1\delta_2\lambda_1\rangle_c+\frac{\langle
    \lambda_1^2\delta_2\rangle_c}{2!}\right]+\cdots \no
\ea
The last expression only shows $C_{NG,3}^{rs}$.  Finally we obtain the
expression for $r_{\delta\delta^s}$, 
\be
\label{eqn:rrs}
r_{\delta\delta^s}(k,u)\simeq
\frac{1+\frac{C^{rs}_{NG,3}(k,u)}{P_{\delta\delta}(k)(1+f\tilde{W}(k)u^2)}}{\sqrt{1+\frac{P_{\theta_S\theta_S}(k)u^4+C_{NG,3}(k,u)+C_G(k,u)}{P_{\delta\delta}(k)(1+f\tilde{W}(k)u^2)^2}}}\ .
\ee
Evaluating $r_{\delta\delta^s}$ requires $P_{\delta\delta}$, $B_3$, $W$ and
$P_{\theta_S\theta_S}$. The first two quantities are observables, as discussed in \S
\ref{sec:RD}. The last two can be inferred from the observed
$P^s_{\delta\delta}$ (\S \ref{subsec:RD}). So there is little uncertainty involved in
predicting $r_{\delta\delta^s}$. This will also be the case for the window function $W^s$
(Eq. \ref{eqn:Ws}) required for velocity reconstruction.

\section{An alternative approach to reconstruct the 3D velocity field}
\label{sec:appendixD}
As shown in \S \ref{sec:3D}, the key to reconstruct the 3D ${\bf v}_\delta$
is to infer the correct window function $W^s$. The approach discussed in \S
\ref{sec:3D} is straightforward. But it relies on the RSD modeling and is hence
susceptible to inaccuracies therein.  Here we propose an alternative to
simultaneously estimate $W^s$  and reconstruct ${\bf v}_\delta$.  It avoids
modeling  the ${\bf v}_\delta$ induced RSD, the most difficult part in the RSD
modeling. So it is less susceptible to uncertainties in the RSD modeling. 

The guideline is that, if we construct $W^s$ and ${\bf
  v}_\delta$ correctly, we can  move particles/galaxies back to their real space
positions. This will eliminate the ${\bf v}_\delta$ induced RSD and hence
reduce anisotropies in the power spectrum after moving (hereafter we denote it as $P^s_{\rm
    move}({\bf k})$). To
further demonstrate this point, let us consider the limit of ${\bf 
  v}_S=0$, 
${\bf v}_B=0$, $\delta_S^s= 0$ and no measurement
noise. Under this limit, a  correct guess of $W^s$ 
will faithfully recover ${\bf v}_\delta$.  Moving the
particles back to their real space positions using this ${\bf v}_\delta$,
  anisotropies in $P_{\rm
    move}({\bf k})$ will be completely eliminated and we
  recover the isotropic real space power spectrum ($P_{\rm move}({\bf
    k})=P(k)$). This suggests that,  by tuning  $W^s$ until the $P_{\rm
    move}({\bf k})$ reaches isotropy, we can recover the 
  correct $W^s$ and hence recontruct the velocity correctly.

The real situation is more complicated, due to the fact that ${\bf v}_S\neq 0$,
${\bf v}_B\neq 0$, $\delta_S^s\neq 0$ and the existence of measurement noise
such as shot noise in the galaxy number distribution. However, none of them is
correlated with the real space density and none of them can cause anisotropic
pattern the same as ${\bf v}_\delta$.  This significantly simplfies the
modeling of $P_{\rm move}$. After we move the particles back according to
${\bf v}_\delta$ reconstructed with the correct $W^s$ (no multiplicative
error), we have 
\ba
P_{\rm move}(k,u)=\int (1+\left\langle
\delta_1\delta_2\right\rangle)D_{\rm move}(k_z,{\bf r})e^{-i{\bf
    k}\cdot{\bf r}}d^3{\bf r}\ .
\ea
Like $P^s$, $P_{\rm move}$ only depends on $k$ and $u$. So we write these
dependences explicitly. Here 
\ba
D_{\rm move}(k_z,{\bf r})&\equiv& \left\langle e^{\lambda_S+\lambda_B-ik_z
  (v^s_{1z,S}-v^s_{2z,S})/H}\right\rangle\ .
\ea
$v^s_S$ is the additive error defined in Eq. \ref{eqn:verror}. 
We also define a $\epsilon_{\rm move}$ through 
\be
1+\epsilon_{\rm move}(k_z,{\bf r})\equiv \frac{D_{\rm move}(k_z,{\bf
    r})}{D_{\rm move}(k_z,r\rightarrow \infty)=D_{\rm move}^{\rm FOG}(k_z)}
\ee

The expression on $D^{\rm FOG}_{\rm move}$ can be further simplified. (1) 
Since ${\bf v}_B$ is uncorrelated with ${\bf v}_S$
and ${\bf v}^s_S$, we have $D_{\rm move}^{\rm FOG}=D_B^{\rm FOG}(k_z)\langle
e^{ik_z \Delta v_S/H}\rangle^2$. Here, $\Delta v_S\equiv
v_{z,S}-v^s_{z,S}$. (2) So far we have neglected shot noise in
the galaxy distribution. Since it does not correlate with other components, it
only causes damping. This effect can be completely described by a damping
function $D^{\rm FOG}_{\rm shot}(ku)$, the same as the case of  ${\bf v}_B$.  For
brevity, we will not  consider this measurement noise hereafter.

The impact of ${\bf v}^s_S$ is harder to deal with, largely due to correlation
between ${\bf v}_S$ and ${\bf v}_S^s$.  To see their
correlation, let us check the limit $k\rightarrow 0$. Now we have
$\delta^s\simeq 
\delta-\nabla_zv_z/H$, the starting point to derive the Kaiser formula. Since
$\delta_s=(\delta-\nabla_zv_{z,\delta}/H)-\nabla_zv_{z,S}/H$, we obtain
$\delta^s_S\simeq -\nabla_zv_{z,S}/H$. Through the relation 
$\theta^s_S({\bf k})=\delta^s_S({\bf k})W^s({\bf k})$, we have ${\bf
  v}_S^s({\bf k})\simeq ({\bf v}_S({\bf k})\cdot \hat{k})u^2W^s({\bf k})
\hat{k}$. Notice that {\it the velocity field ${\bf v}_S^s$ is statistically
anisotropic}. Under this limit, ${\bf v}^s_S$ is completely correlated with ${\bf
  v}^s$.

Due to this correlation,  we can not
calculate  $\langle \cdots \rangle^2$ seperately for $v_{z,S}$ and
$v^s_{z,S}$.   Neverthless, using the relation 
\ba
1+\epsilon_{\rm move}(k_z,{\bf r})=e^{k_z^2\langle
    \Delta v_{1S}\Delta v_{2S}\rangle/H^2+O(\Delta v_S^4)}\ , 
\ea
we obtain
\ba
\label{eqn:move}
P_{\rm move}(k,u)&=&\left[P_{\delta\delta}(k)+P_{\Delta v\Delta v}(k,u)
  u^4+\cdots \right]\no \\
&&\times D^{\rm
  FOG}_{\rm move}(k_z)\ . 
\ea
Here, $P_{\Delta v\Delta v}(k,u)$ is the power spectrum of $\Delta {\bf v}_S\equiv
{\bf v}_S-{\bf v}^s_S$.  Due to the intrinsically anisotropic ${\bf v}^s_S$, $P_{\Delta
  v\Delta v}({\bf k})$ is also anisotropic and depends on both $k$ and $u$.  At large scale limit,  
\ba
P_{\Delta v\Delta v}({\bf
  k})&\simeq& P_{\theta_S\theta_S}(k)(1-u^2W^s({\bf k}))^2\no \\
&\simeq&
P_{\theta_S\theta_S}(k)\frac{1}{(1+f\tilde{W}u^2)^2}\ .
\ea 
This results shows a generic property $P_{\Delta v\Delta  v}(k,u)/P_{\delta\delta}(k)\rightarrow 0$
when $k\rightarrow 0$, since both  ${\bf v}_S$ and ${\bf v}_S^s$
vanish at large scales. We propose that, by tuning $W^s$ such that $P_{\rm move}$ follows a form like
Eq. \ref{eqn:move},  we could obtain the correct $W^s$ and
hence the correct ${\bf v}_\delta$ field.

The above result is obtained in the large scale limit. The situation
beyond this limit is too complicated to discuss analytically and will be postponed
for future study.  For the same reason,  we still lack of a rigorous
mathematical proof nor numerical verification for the above proposal. However,
given its potential in reconstructing the 3D peculiar velocity in a less model
dependent way, we hope
to explore this possibility in future works. 
\bibliography{mybib}

\end{document}